\documentclass[epj,referee]{svjour}

\usepackage{latexsym}
\usepackage{graphics}

\usepackage{amsfonts}
\usepackage{amssymb}
\usepackage{subfigure}
\usepackage{textcomp}
\usepackage{gensymb}
\usepackage[english]{babel}
\usepackage[T1]{fontenc}
\usepackage{setspace}
\usepackage{color}
\usepackage{siunitx}
\usepackage{isotope}
\usepackage{here}
\usepackage{afterpage}
\begin{document}
%
%%\title{``Ping-pong'' with ultracold neutrons to understand transport properties.}
\title{Ultracold neutron storage and transport at the PSI UCN source}
\author{
%Me\inst{1}, 
%him\inst{2}, 
%and the others\inst{1,2}
%
G.~Bison\inst{1},
M.~Daum\inst{1},
%L.~G\"oltl\inst{1,2},
K.~Kirch\inst{1,2},
%S.~Komposch\inst{1,2},
B.~Lauss\inst{1} \thanks{Corresponding authors:  \\geza.zsigmond@psi.ch (G. Zsigmond) \\bernhard.lauss@psi.ch (B. Lauss)},
%I.~Rienäcker\inst{1,2},
D.~Ries\inst{3},
P.~Schmidt-Wellenburg\inst{1},
G.~Zsigmond\inst{1} \footnotemark[1]
}
\institute{$^1$Laboratory for Particle Physics,
Paul Scherrer Institut, CH-5232 Villigen-PSI, Switzerland \\
$^2$Institute for Particle Physics, ETH Z\"urich, Z\"urich, Switzerland \\
$^3$Department of Chemistry - TRIGA site, Johannes Gutenberg University Mainz, 55128 Mainz, Germany \\
}
%\date{Received: date / Revised version: date}
\date{\today}% It is always \today, today,
\abstract{
Efficient neutron transport is a key ingredient to the performance of ultracold neutron (UCN) sources, important to meeting the challenges placed by high precision  fundamental physics experiments. At the Paul Scherrer Institute's UCN source we have been continuously improving our understanding of the UCN source parameters by performing a series of studies to characterize neutron production and moderation, and UCN production, extraction, and transport efficiency to the beamport. 
The present study on the absolute UCN transport efficiency completes our previous publications. We report on complementary measurements, namely one on the height-dependent UCN density and a second on the transmission of a calibrated quantity of UCN over a $\sim$16 m long UCN guide section connecting one beamport via the source storage vessel to another beamport. 
These allow us quantifying and optimizing the performance of the guide system based on extensive Monte Carlo simulations.
%These allow us to quantify the performance of the guide system and to optimize the vertical position of UCN storage experiments. 
%These measurements are compared to detailed Monte Carlo simulations of the neutron  optics. In general, there is a very good agreement between measurements and simulations, which enables us to constrain the parameters of the model including the UCN energy spectrum emerging from the solid deuterium (sD$_2$) moderator, consistent with an imperfect material structure. 
}
%end of abstract
 % end of PACS codes
\authorrunning{G.~Bison et al.}
\titlerunning{Ultracold neutron storage and transport at the PSI UCN source}
\maketitle
\onecolumn
% remove for double column version
%  forced ending of subfile
%%  \endinput
%%%%%%%%%%%%%%%%%%%%%%%%%%%%%%%%%%%
\section{Introduction}
%%% \section{Introduction}
%%% \input{InputTex/introduction}

Ultracold neutrons (UCNs) are neutrons with kinetic energies below about 300\,neV, equivalent to temperatures below 3\,mK. They are key probes in particle physics experiments at the low-energy frontier.
UCNs are reflected at any angle of incidence from certain materials like steel, beryllium, nickel, diamond-like carbon (DLC) or alloys like 
nickel-molybdenum (NiMo), since they cannot penetrate a sufficiently thick nuclear potential barrier of the material which is higher than their kinetic energy. 
%The higher the potential barrier of the material the higher is the kinetic energy range of the UCNs.
%
Hence, UCNs can be stored in bottles made from or coated with these materials and
%The exact range of kinetic energy up to which neutrons can be reflected depends on the height of the potential barrier, which in turn is dependent on the atomic density and the coherent scattering length of the material.
%There is still a small probability that during reflection the neutrons penetrate into the coating (to a depth comparable to their wavelength) and a small fraction of them is absorbed in or up-scattered by the nuclei. This sets a limit to the storage time, but even so, they 
can be observed for hundreds of seconds~\cite{Golub1991}.
This makes it possible to precisely examine the intrinsic properties of the neutron, 
and to search for physics beyond the Standard Model of particle physics (BSM).
The most prominent example is the search for a  permanent electric dipole moment of the 
neutron (nEDM)~\cite{Baker2006,Baker2011,SerebrovEDM2015,Pendlebury2015,Ito2018,Ahmed2019,Wurm2019,Martin2020,Abel2020}.
The best achieved sensitivity for such experiments is limited by counting statistics, and efforts are made worldwide to develop new 
UCN sources to provide higher  intensities~\cite{Bison2017,Leung2019source,Schreyer2020}.
Besides a high source yield, a highly optimized neutron transport to beamports and experiments is a key ingredient to the performance of UCN sources. 

The high intensity UCN source at the Paul Scherrer Institute 
%(PSI)~\cite{Bison2017,Anghel2009,Lauss2011,Lauss2012,Becker2015,Blau2016,Goeltl2012,Ries2016,Bison2020,Lauss2021scipost} 
(PSI)~\cite{Bison2017,Lauss2014,Becker2015,Bison2020,Lauss2021} has been in operation since 2011. 
The main UCN-related parts of the source are shown
in the center of Fig.~\ref{fig:ping-pong-setup}.
The operation scheme follows this sequence:
Every 300\,s the full proton beam \textendash\, 590\,MeV and up to 2.4\,mA \textendash\,  is directed onto the spallation target for up to 8\,s
(label TAR in~Fig.\ref{fig:ping-pong-setup}.)~\cite{Wohlmuther2006}.
(The measurements described here were performed at a time when only shorter pulse lengths were allowed for regulatory reasons.)
Fast neutrons produced in the lead target material are thermalized at room temperature in the surrounding heavy water (D$_2$O).
A fraction of them is further thermalized to the cold neutron regime in solid deuterium (sD$_2$) at 5\,K, and in a last step down-scattered via superthermal conversion into the ultracold energy range.
UCNs can then propagate from the sD$_2$-vessel upwards  into vacuum through a vertical guide and fill the source storage vessel (SV).
The two flaps on the bottom of the vessel are closed at the end of the proton pulse 
and only reopened again shortly before the next beam pulse.
From the storage vessel of the source the UCNs can propagate via two guides connected to the bottom of the SV  (dubbed ``West-1'' and ``South'')
to the experimental areas. Similarly, with lower intensity, UCNs reach a guide connected to the top of the SV (``West-2''), which is used for test measurements. 

The UCN intensity at the PSI source was considerably increased over the last years. 
At the same time, we have been improving our understanding of the UCN source parameters. 
%von Klaus:
In a series of studies, we covered neutron production and moderation, UCN production and extraction out of the sD$_2$ and the moderator vessel, and aspects of the UCN optics~\cite{Lauss2014,Becker2015,Bison2020,Atchison2009a,Atchison2011,Anghel2018}.
The results of our measurements are overall well understood and reproduced with detailed simulation models of the UCN source. In the present work, we elaborate on the energy-dependent, absolute UCN transport efficiency of the UCN guide system, and on the UCN energy spectrum delivered at the beamports.
%This was achieved with  a series of studies covering:
%\begin{itemize}
%\item Neutron production and moderation, already experimentally tested~\cite{Becker2015}. A very good agreement between simulation and measurement was found.
%\item UCN production and extraction from the sD$_2$ and the moderator vessel. There a further increase of the UCN yield seems possible on the basis of improved cryogenic operation parameters (to be published).
%\item Energy-dependent UCN transport efficiency to the beamports on which we will elaborate in this paper.
%\end{itemize}

UCN transport characteristics of the individual neutron guides were measured prior to mounting, and the results reported in Ref.~\cite{Blau2016}. 
A previous characterization of the UCN optics system at the PSI source after commissioning based on Monte Carlo simulations was published in Ref.~\cite{Bison2020}. In the present study we report on further complementary measurements of the UCN density in a storage bottle at varying heights above the beam axis. A study of the UCN transport  using the entire guide system of the source and a calibrated UCN quantity for absolute efficiencies supplements this report. 
The results allow positioning of experiments at heights optimized for UCN density. 
The results also improve our knowledge of the actual neutron optics parameters of the UCN guides and the source storage vessel after installation. 
By limiting the energy range of the neutrons to the UCN regime by the storage method, thus eliminating very cold neutrons, we could further constrain the simulation model parameters. We additionally present a realistic profile of the initial energy spectrum of neutrons exiting the sD$_2$ moderator. 
In our previous work~\cite{Bison2020}, this was assumed to be linear, characteristic of an ideal UCN moderator material, according to Eq. (3.9) in~\cite{Golub1991}. However, measurements and simulations indicate~\cite{Anghel2018} that, along with material cracks in the bulk, a frost layer on the sD$_2$ surface causes energy dependent losses of UCNs, due to back-reflection from crystallites. Thus in the simulations described in this work, we parametrize and fit the shape of the spectrum.
In-situ measurements of the UCN transport also include all eventual inefficiencies like coating deficiencies, gaps between guide sections and shutters, or imperfect mounting. These additional losses are considered in the simulation model using effective parameters. 

The obtained results provide fundamental input to better analyze the UCN source performance, help to identify sections for further improvements, and allow for more precise estimations of the statistical sensitivity of experiments.
In Sec.~\ref{Sec:experiment}, experimental setups and results will be presented, which are subsequently compared to detailed Monte Carlo simulations in Sec.~\ref{Sec:simulation}.

%\clearpage

\section{Transmission experiments with stored UCNs}
\label{Sec:experiment}

In the following we describe the two main experiments: 
(a) UCN transmission from beamport West-1 to beamport South, allowing to constrain the loss and diffuse reflection parameters of the guide coatings, and 
(b) UCN storage at different heights above beamport level, 
aiming to find the vertical position maximizing UCN density, and helping to constrain the shape of the energy spectrum of the UCN exiting the sD$_2$ moderator material. Comparison of the second experiment with simulation helps to take into consideration the energy dependence of losses in experiment (a). Vice versa, the results from experiment (a)  will influence the energy profile of the stored neutrons in experiment (b). Consequently, the simulation analysis has to take into account this inter-dependence of the results from the two experiments, as it will be detailed in Sec.~\ref{Sec:simulation}.

%%%%%%%%%%%%%%%%%%%%%%%%%%%%%%%%%%%
%%\section{The Ping Pong Experiment}

\subsection{Beamport-to-beamport transmission}
\label{pingpongexperiment}

\subsubsection{Motivation and principle of the measurement}
\label{motivation}

The  measurements presented in this section test the neutron optics above the central storage vessel flaps.
The idea of the measurement, with the involved parts illustrated in Fig.~\ref{fig:ping-pong-setup}, is the following: 
UCNs are produced and filled into the source
storage vessel (SV in Fig.~\ref{fig:ping-pong-setup}). 
The neutron guide shutters (g)  towards both beamports are always open and not used in this measurement. UCNs are also filled directly into an external storage vessel (b), where they are stored for \SI{130}{s}. 
During this time the  storage vessel of the source 
and the guide system are cleaned of UCNs by opening
the central storage vessel flaps (f), allowing the UCNs to fall back towards the solid D$_2$ where they are lost. 
After this procedure, the central flaps are closed again
to later allow the UCNs to traverse the storage vessel.
Subsequently, the UCNs are released from the external storage vessel by opening shutter 1, permitting back-propagation towards the storage vessel of the source, and from there  into the guide South. 
A detector mounted at beamport South detects the UCNs which traversed both 
guides and the source storage vessel.
As the UCNs are sent first from the storage vessel of the source to beamport West-1, and then from there to beamport South,
the measurement is dubbed UCN ``ping-pong''.

In order to quantify how many UCNs started towards the  beamport South after their release from the external storage vessel at beamport West-1, a reference measurement was necessary.  In this reference measurement the UCNs were released by opening shutter 2, on the other side of the external chamber, towards detector West-1. 
By using identical filling and storage times as in the ping-pong procedure, and by performing the measurements within a short time to assure UCN source stability, we made sure that the correct initial number of UCNs was obtained.

The transmission efficiency of the UCNs from one beamport to the other beamport then depends on all cumulative losses in the guides, windows, storage vessel, and possible slits between the components.
Although all UCN guides were tested prior to installation and commissioning of the UCN source~\cite{Goeltl2012}, %,Bison2020,Bison2016},
this in-situ measurement provides additional information on unavoidable imperfections in the assembly as well as aging or contamination since the ex-situ 
measurements.
First test measurements were reported in~\cite{Goeltl2012}. 
These were repeated after improvements using (i) an automatic timing sequence for the central storage vessel flaps, (ii) higher time resolution in the UCN detection. 
Additionally, the measurements reported here include operation of the polarizing magnet (d in Fig.~\ref{fig:ping-pong-setup}) at beamport South at various field strengths. Thus we obtained the beamport-to-beamport transmissions for different UCN energy spectra, and can directly compare these to the values extracted from Monte Carlo simulations.
%%%%%%%%%%%%%%%%%%%%%%%%%%%%%%%%%%%%%%%%%%%%%%%%%%%%%%%%%%%%%%%%%%%%%%%%%%%%%%%%%%%

\subsubsection{Setup}

\begin{figure*}[htb]
\begin{center}
\resizebox{0.99\textwidth}{!}{\includegraphics{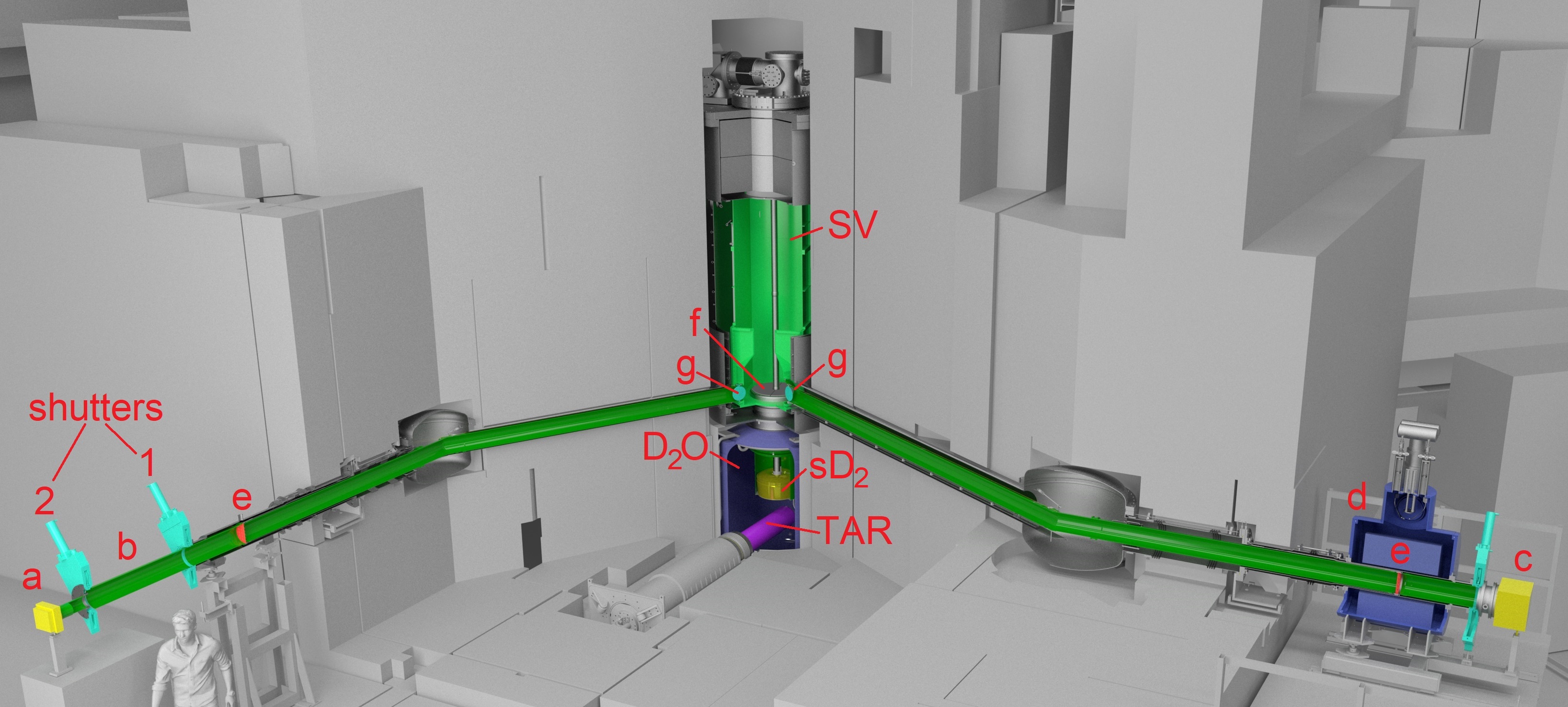}
}
\caption [Setup for the UCN ping-pong measurement] {
CAD drawing of the most important parts of the UCN source 
installation used for the measurements described: 
(a) small Cascade detector connected to the storage
bottle at beamport West-1 with 
a \SI{300}{\milli\meter} long I.D. \SI{100}{\milli\meter} NiMo-coated guide; 
(b) storage bottle made of a \SI{1}{\meter} long
glass UCN guide coated with \isotope{NiMo} and 
two DLC-coated vacuum shutters (1, 2),
(c) big Cascade detector mounted at beamport South with a 
\SI{150}{\milli\meter} long I.D. \SI{180}{\milli\meter} NiMo-coated guide, 
(d) polarizing magnet,
(e) AlMg3 vacuum safety windows,
(f) central storage vessel flaps,
(g) neutron guide shutters,
(TAR) lead spallation target,
(D$_2$O) heavy water container,
(sD$_2$) solid deuterium moderator vessel,
(SV) UCN source storage vessel. The biological shielding made of iron and concrete blocks is depicted in grey.
The UCN guide on top of the source storage vessel towards beamport West-2 is omitted, details can be found in Ref.~\cite{Bison2020}.
}
\label{fig:ping-pong-setup}
\end{center}
\end{figure*}

A \SI{25}{\liter} 
external storage vessel made of a \SI{1}{\meter} long piece of 
 UCN guide  (NiMo coated glass tube I.D. \SI{180}{\milli\meter}) and two 
shutters (1,2) coated with DLC were mounted at beamport West-1. 
UCNs could be emptied in both directions, 
either 
towards the UCN storage vessel of the source via shutter 1
or
towards a small  (10\,cm $\times$ 10\,cm active surface) Cascade UCN detector~\cite{Cascade2013} via shutter 2.
A CAD drawing of the setup is shown in
Fig.~\ref{fig:ping-pong-setup}. The total length of the UCN guides from the source storage vessel to beamport West-1 is  7090\,mm, and to beamport South is 8618\,mm, with the length being the sum of all parts according to construction drawings.
The shutters were actuated by pressurized air and took approximately 
\SI{1}{\second} to open or close, as described in~\cite{Goeltl2012}.%,Bison2020}.

% 
% At beam port West-1 a UCN detector --- a small Cascade counter, attached
% to the \SI{25}{\liter} storage vessel via a \SI{30}{\centi\meter} long 
% UCN guide made from acrylic glass, ID=\SI{100}{\milli\meter}, coated with NiMo
% --- can measure
% the amount of UCN after a certain storage time.
% Thus, assuming
% identical emptying characteristics through each of the two shutters,
% a known amount of UCN can also be sent towards a big Cascade detector mounted
% at the beam port on guide South.

At beamport South, a big (20\,cm $\times$ 20\,cm active surface) Cascade detector~\cite{Cascade2013}  
was mounted right outside
the 
%``Strahlwegventil'' 
beamport shutter, i.e. behind the \SI{5}{\tesla} polarizing magnet.
Measurements without and with magnetic fields of \SI{1.5}{},
\SI{2}{}, \SI{3}{}, and \SI{5}{\tesla} were conducted
aiming at a certain velocity discrimination of the arriving UCNs, and depending on polarization state.

%%%%%%%%%%%%%%%%%%%%%%%%%%%%%%%%%%%%%%%%%%%%%%%%%%%%%%%%%%%%%%%%%%%%%%%%%%%%%%%%%%%

\subsubsection{Measurement and timing}

Two separate control systems were used. 
The central storage vessel flaps were operated by the UCN source control system.
As under normal operation conditions, the flaps were closed on a trigger
from the accelerator system at a constant time before 
the end of the proton beam pulse. 
Instead of keeping the flaps closed until shortly before the next
proton beam pulse, as in standard operation, 
the flaps were opened after shutter 1 was closed in order to drain the source from remaining UCNs. 
The flaps were closed again another \SI{110}{\second} later,  before reopening the external storage vessel shutter. 
Once the measured UCN count rate was at background level the flaps were opened to be ready to receive the next proton beam pulse.

The shutter timing system, as described in~\cite{Goeltl2012} was triggered by the falling edge of the accelerator signal occurring at the end of the proton beam pulse.
Both UCN detectors were triggered simultaneously on the accelerator signal 
%`WWK' 
generated before the beam pulse, 
%the WWK signal by approximately \SI{8}{\second}, 
which resulted in the time-base
of the Cascade data files starting at approximately \SI{7.7}{\second} before the
beginning of the proton beam pulse. 
They then ran on their internal clocks.
UCN measurements at beamports South (ping-pong) and West-1 (for reference)  were done in multiple groups of typically five repetitions in order
to avoid systematic effects due to drifts of the source performance~\cite{Anghel2018}.

\begin{figure*}[htb]
\begin{center}
\resizebox{0.70\textwidth}{!}{\includegraphics{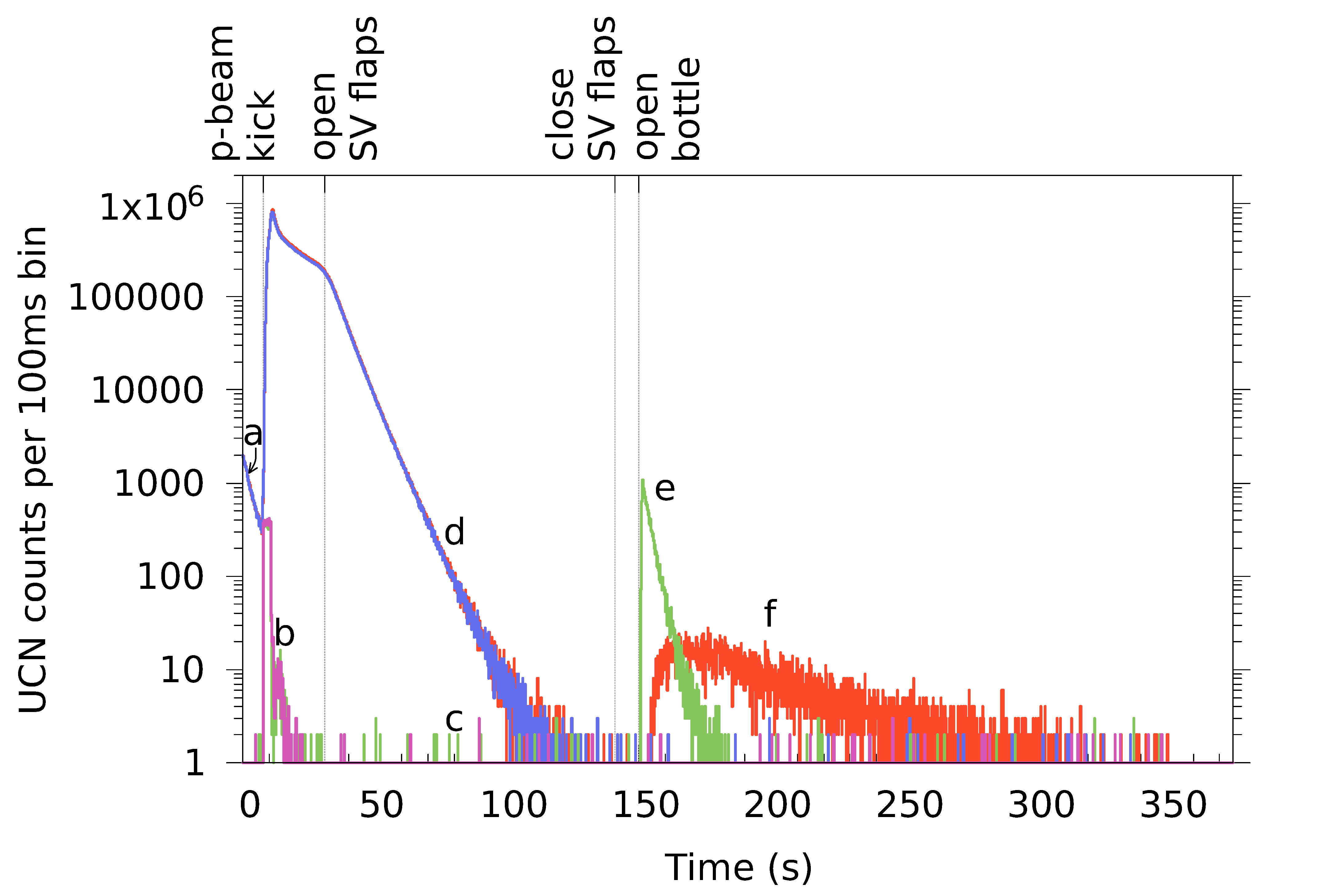}
}
\caption [UCN ping-pong results]
{Sum of UCN counts of 21 measurements opening the external storage bottle 
towards the UCN source and 20 measurements opening towards the external detector
at beamport West-1. The counts detected by the two detectors (a, c in Fig.~\ref{fig:ping-pong-setup}) are overlayed in the time spectrum.
a) Few UCNs produced in the \SI{5}{\milli\second} long pilot beam kick.
b) Neutrons passing through closed shutters into the detector at beamport West-1 during the proton beam pulse and shortly after it (magenta).
c) Background level in detector West-1.
d) Quickly decreasing UCN count rate at beamport South while storage vessel 
flaps are open (blue).
e) UCN counts in detector West-1 (reference measurement) after opening shutter 2 (green). 
f) UCN counts measured at beamport South (ping-pong) after opening shutter 1 (red).
The polarizing magnet was off.}
\label{fig:ping-pong-data}
\end{center}
\end{figure*}

\begin{figure*}[htb]
\begin{center}
\resizebox{0.70\textwidth}{!}{\includegraphics{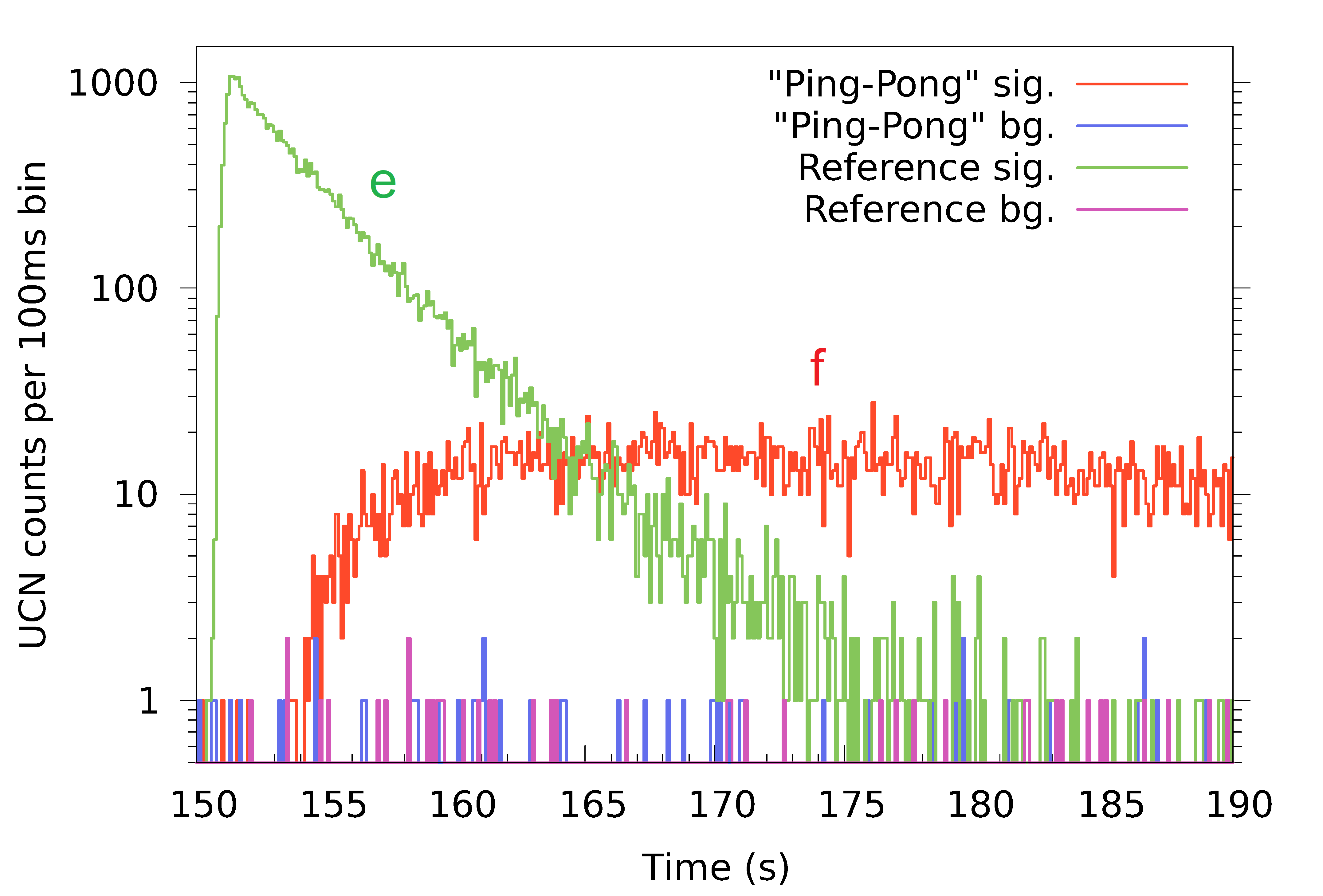}}
\caption [UCN ping-pong results, zoom]
{Zoom into the region of interest of Fig.~\ref{fig:ping-pong-data}. 
The first \SI{40}{\second} after opening
the shutter of the external storage vessel are shown. 
The arrival time difference
of about \SI{4}{\second} between the first UCNs detected in the directly connected
detector at beamport West-1 (green line (e))
and the first UCNs detected after traversing
the source 
at beamport South (red line (f))
is well visible. 
Background counts in one detector were measured when UCNs
were emptied towards the other side. 
The background level is comparable to the one 
measured when the UCN source is not operating.
}
\label{fig:ping-pong-data-zoom}
\end{center}
\end{figure*}

The time sequence of the measurement was the following (all neutron guide shutters were open all the time): 
\begin{itemize}
\item  $t$ = 0, start of the detector data file. 
\item  $t$ = 7.7\,s,   start of the 3\,s proton beam pulse (longer pulses were not allowed at the time of the here described measurements due to regulatory limitations).
\item  $t$ = 10.7\,s,  end of the proton pulse.
\item  $t$ $\simeq$ 10.7\,s, source vessel flaps close approximately at the end of the proton pulse.
\item 
during and after the proton beam pulse the external storage vessel is filled for \SI{8.6}{\second} with shutter 1 open and shutter 2  closed 
(this filling time was optimized beforehand for a maximum number of stored UCN). 
\item
$t$ = 19.3\,s, shutter 1 closes.% (the storage time constant of the source storage vessel in this state with guide South and West-1 open is \SI{22.1}{\second}). 
\item
$t$ = 20.7\,s, i.e.  \SI{1.4}{\second} later, the main flaps 
of the UCN storage vessel open and for \SI{110}{\second} the UCNs can fall down towards the solid deuterium vessel where they are absorbed. 
Hence, the storage vessel and connected guides are quickly emptied.
The UCN count rate in detector South drops rapidly after opening the flaps 
and reaches the background level of $\ll$ 1\,Hz
after about \SI{100}{\second}. 
The storage time constant with open flaps is approximately \SI{6.0}{\second}. 
\item
$t$ = 130.7 s, the storage vessel flaps close again. 
\item
$t$ = 149.3 s, after a storage   time of \SI{130}{\second}, one
shutter of the external storage vessel opens, 
releasing the stored UCNs. 
When shutter 2 opens, a reference measurement of the stored UCNs is done. 
When shutter 1 opens, a transport measurement to beamport South is done.
The detector behind the closed shutter 2 measures the background simultaneously.
After traversing the source UCN guides and the
source storage vessel, a fraction of the UCNs 
reaches the detector at beamport South. 
The observed number of UCN counts with respect to the
time after the beam pulse is shown 
 in Figs.~\ref{fig:ping-pong-data} and~\ref{fig:ping-pong-data-zoom} for both beamports.
One can observe that the arrival peak of the UCNs after opening the external storage bottle 
%either 
%towards the directly attached detector or back towards the source 
is well separated from the UCN production peak, hence the cleaning of the storage vessel worked well. 
\item 
$t$ = 330 s, all shutters are reset into the initial position, shutter 1 open, shutter 2 closed, storage vessel flaps open.
\item
$t$ = 360 s, next proton pulse.
\end{itemize}

Fig.~\ref{fig:ping-pong-data-zoom}, a zoom of Fig.~\ref{fig:ping-pong-data}, shows the UCN counts for times larger than 150\,s, i.e. after the release of the UCNs towards beamport South.
One can see that the storage time of \SI{130}{\second} was long enough such
that all UCNs were drained from the source storage vessel before the 
external vessel was opened again.
One can also see the time difference of about 3.5\,s between the UCNs arriving in detectors South (e) and West-1 (f) in the measurements, when emptying the external storage volume to the respective side. This corresponds to the traversing time through the source system. The different profiles of the UCN counts are caused by the different UCN path lengths,  together with the time-of-flight spread from the spectrum, the geometry and wall-scattering during transport through the guides and the passage through the source storage vessel. The slope in the count rate of the detector at beamport West-1  reflects the emptying time of the external storage vessel. 
All measurements are done with a proton beam pulse repetition 
time of \SI{360}{\second} to ensure that all UCNs are drained from the entire system.

The big and small Cascade detectors have different detection efficiencies, due to their difference in size, different connection to the UCN guide, and different aluminum foils serving as entrance window, which can have different transmissions~\cite{Goeltl2012}.
This resulted in different detection efficiencies, depending also on the time of UCN storage in the external storage bottle.

With detectors directly attached to the external storage bottle at beamport West-1, we measured their relative detection efficiencies, $\epsilon$, in similar conditions, i.e. only separated by the time it took to unmount one and mount the other detector. We found a relative detection efficiency of the small to the big detector of 
$\epsilon$ = 0.53$\pm$0.01 after 130s storage in the external storage vessel. It is important to compare after the same storage time as in the measurement, as the UCN velocity spectrum gets slower with longer storage times and the detection efficiency depends on velocity.

%%%%%%%%%%%%%%%%%%%%%%%%%%%%%%%%%%%%%%%%%%%%%%%%%%%%%%%%%%%%%%%%%%%%%%%%%%%%%%%%%%%

\subsubsection{Results and discussion}
\label{Sec:pingpong-results}

The different detection efficiencies were taken into account 
when %calculating the transmission of UCNs from the external storage vessel through the source guides and storage vessel to the detector at beamport South.
we calculate the corrected fraction, $R$ of UCNs as  
\begin{equation}
R =\frac{N_\text{South}}{N_\text{West-1}} \times \epsilon, 
\end{equation}
where the counts $N_\text{South}$ (ping-pong) and $N_\text{West-1}$ (reference) are evaluated for the entire time interval when shutter 2 is open, i.e. 149\,s - 330\,s., 
and $\epsilon$ is the measured ratio of detector efficiencies.

The corrected fraction of UCNs reaching the detector at beamport South is   
0.135$\pm$0.011 
with the polarizing magnet off, and
%of \SI{0.070+-0.001}{} 
0.093$\pm$0.006 for a field strength of \SI{5}{\tesla}, where the errors stem from the standard deviation of counts per pulse and reflect the small fluctuations in the yield of the UCN source. 
Results for all magnetic field settings as well as the number of repetitions for both
ping-pong signal and reference measurements are given in Table~\ref{tab:PinpongResTable}. In the reference measurements the detector at beamport South measured the  background for the transmission measurement, and vice versa.

\begin{table*}[htb]
\begin{center}
\begin{tabular}{|c|c|c|c|c|c|}
\hline
B (T)  & fraction   & corrected fraction & signal rep. & reference rep. &  simulated fraction    \\ \hline
0      & 0.254$\pm$0.022  & 0.135$\pm$0.011       & 21        & 20   &  0.150$\pm$0.006   \\
1.5    & 0.186$\pm$0.010  & 0.099$\pm$0.006     & 15        & 6    &  0.084$\pm$0.004    \\
2      & 0.167$\pm$0.010  & 0.089$\pm$0.006     & 14        & 9    &  0.087$\pm$0.005 	  \\
3      & 0.166$\pm$0.019  & 0.088$\pm$0.011       & 17        & 8    &  0.086$\pm$0.003 	  \\
5      & 0.176$\pm$0.012  & 0.093$\pm$0.006       & 28        & 11   &  0.091$\pm$0.005    \\
\hline
\end{tabular}
\caption[UCN ping-pong results]
{Results for the ping-pong measurements. 
The fraction of UCN counts
at beamport South with respect to UCN counts at beamport West-1 and the 
same fraction corrected for detector efficiencies are given. Standard deviation errors are used in order to cover small non-statistical fluctuations.% probably due to a later resolved small timing issue in the UCN source control.
The number of repetitions (rep.)
is given separately for signal measurements (opening shutter 1 towards beamport South)
and reference measurements (opening shutter 2 towards the detector at beamport West-1).
Values for the simulated fraction are explained in Sec.~\ref{Sec:simulation}.
}
\label{tab:PinpongResTable}
\end{center}
\end{table*}

The UCN transmission of the safety window in the center of the magnet decreases 
with higher field-values
as expected, as UCNs with the ``right'' spin state are accelerated
by the magnetic field ('high field seekers') and pass, 
while UCNs with the ``wrong'' spin
state ('low field seekers') are decelerated, and can only pass if their kinetic energy
is higher than the repelling potential of \SI{60.3}{\nano\electronvolt/\tesla}.
The polarizing magnet can fully polarize the UCN beam, because the maximum magnetic field-strength of \SI{5}{\tesla} results in a
potential barrier of 300\,neV for the ``wrong'' spin state that is higher than the Fermi potential of all guide walls (about  220\,neV).
However, the effective transmission rate of the magnet, taking into account
the vacuum safety window made of a \SI{100}{\micro\meter} thick AlMg3 foil, mounted at the point of highest magnetic field in the center of the magnet, does not decrease to~0.5. 
This is because high-field seeking UCNs are accelerated and hence have a higher transmission through the foil.

%See also Chapter magnet section 
%~\ref{magnetsection} 
%for more details on UCN transmission through
%the magnet with respect to the magnetic field.

\begin{figure*}[htb]
\begin{center}
\resizebox{0.50\textwidth}{!}{\includegraphics{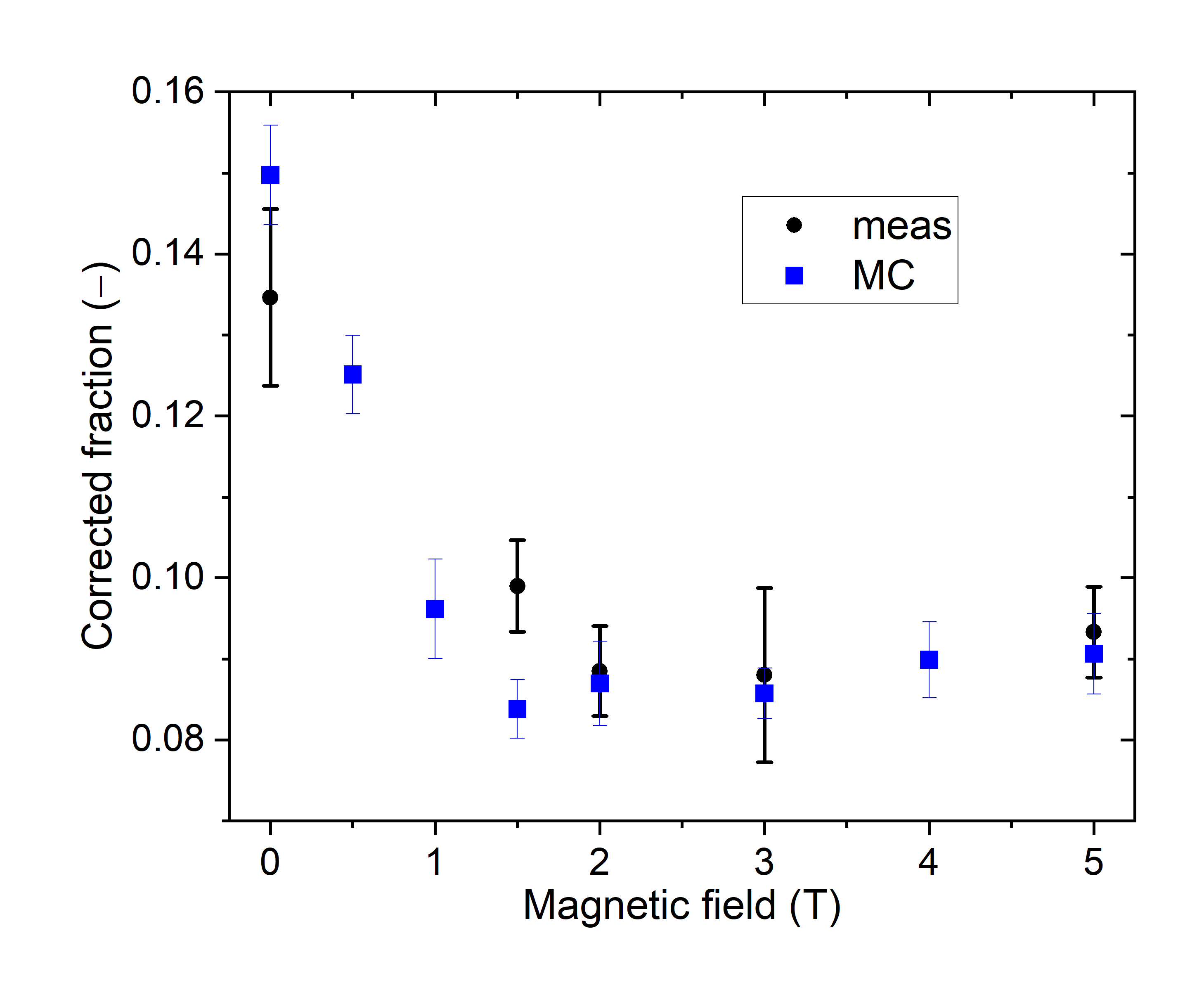}}
\caption [UCN ping-pong transmission vs magnetic field.]
{
Corrected fraction of UCNs detected at beamport South for different  magnetic-field values of the polarizing magnet at beamline South, 
between 0\,T and 5\,T. Filled squares are from the full simulation (see Sec.~\ref{Sec:simulation}).
}
\label{fig:PingPongFractionSouthWestvsSCfieldPow2.7}
\end{center}
\end{figure*}

The measured and corrected fractions of UCNs with respect to the magnetic field
strength in the polarizing magnet are shown in
Fig.~\ref{fig:PingPongFractionSouthWestvsSCfieldPow2.7}. The observed magnetic field dependence is a result of two effects. On the one hand, with increasing magnetic field,  less UCNs with the ``wrong'' spin can pass and the corrected fraction decreases. On the other hand, up to 1\,T fields, an increasing number of ``good'' spin UCNs receives enough energy boost to be transmitted through the potential barrier of the AlMg3 safety foil (54\,neV). For UCNs with kinetic energies below the optical potential of the bottle  (220\,neV), 3.7\,T is sufficient to fully polarize them. Due to 130\,s storage, the average kinetic energy of the UCNs is well below the optical potential, thus the polarization is completed at  magnet fields lower than 3.7\,T, as visible in Fig.~\ref{fig:PingPongFractionSouthWestvsSCfieldPow2.7}.

Fitting the time distribution of the UCN counts from the external storage bottle of the reference measurement yields an emptying time constant of 3.08$\pm$0.02\,s. 
Fitting
the rate of UCNs detected at beamport South after traversing the source,
excluding the rising edge, yields a time constant of 
35.0$\pm$1.4\,s. 
This is compatible with measurements of the
emptying time constant of the UCN source storage vessel with
both beamports South and West-1 opened, at times larger than \SI{150}{\second}
after the proton beam pulse, which is 35.8$\pm$0.2\,s.

As additional input for the simulation we also measured the UCN transmission through the AlMg3 foils from the same batch as the guide safety windows (marked `e' in Fig.~\ref{fig:ping-pong-setup})  installed in a similar geometry in the same setup after \SI{130}{\second} of storage. The result was  0.70$\pm$0.03 for the integral kinetic energy spectrum.

%%%%%%%%%%%%%%%%%%%%%%%%%%%%%%%%%%%
%%\section{The Steel Bottle vs Heights Experiment}
\subsection{UCN storage bottle measurements at various heights}
\label{SteelBottleHeights}

\subsubsection{Motivation and measurement principle}

In a second experiment we conducted measurements of stored UCNs in a stainless steel UCN storage bottle with fast shutters,  which was built to compare UCN densities at various UCN sources~\cite{Bison2017} and hence was transportable. 
We performed measurements to determine the UCN density at different heights above beamport West-1. One goal was to find the  optimal height with maximal UCN density. Another goal was to derive information about the UCN spectrum itself. The storage bottle, as ``standard volume'', was characterized in detail in~\cite{Bison2017,Bison2016}, and hereafter will be referred to as ``standard storage bottle''.
The idea behind varying the height of the standard storage bottle is  
to adapt the UCN energy spectrum delivered by the UCN source to its storage properties.
At beamport West-1, the energy spectrum is cut off at 
\SI{54}{\nano\electronvolt} at the lower end by the AlMg3 vacuum safety window
located approximately \SI{530}{\milli\meter} upstream of the beamport shutter.
As the maximal storable UCN energy is given by the neutron optical potential of the 
surface material of the standard storage bottle, such a lower cutoff effectively limits the 
phase space of UCNs that can contribute to the UCN density.
A rise in height shifts all UCN energies to lower values due to gravity, 102.5 neV per meter rise, and thus can compensate for this lower energy cutoff.
If UCN losses in the guides between the beamport and the elevated standard storage
bottle are low, the net density inside the latter can be increased.

\subsubsection{Setup}

\begin{figure*}[htb]
\begin{center}
\resizebox{0.70\textwidth}{!}{\includegraphics{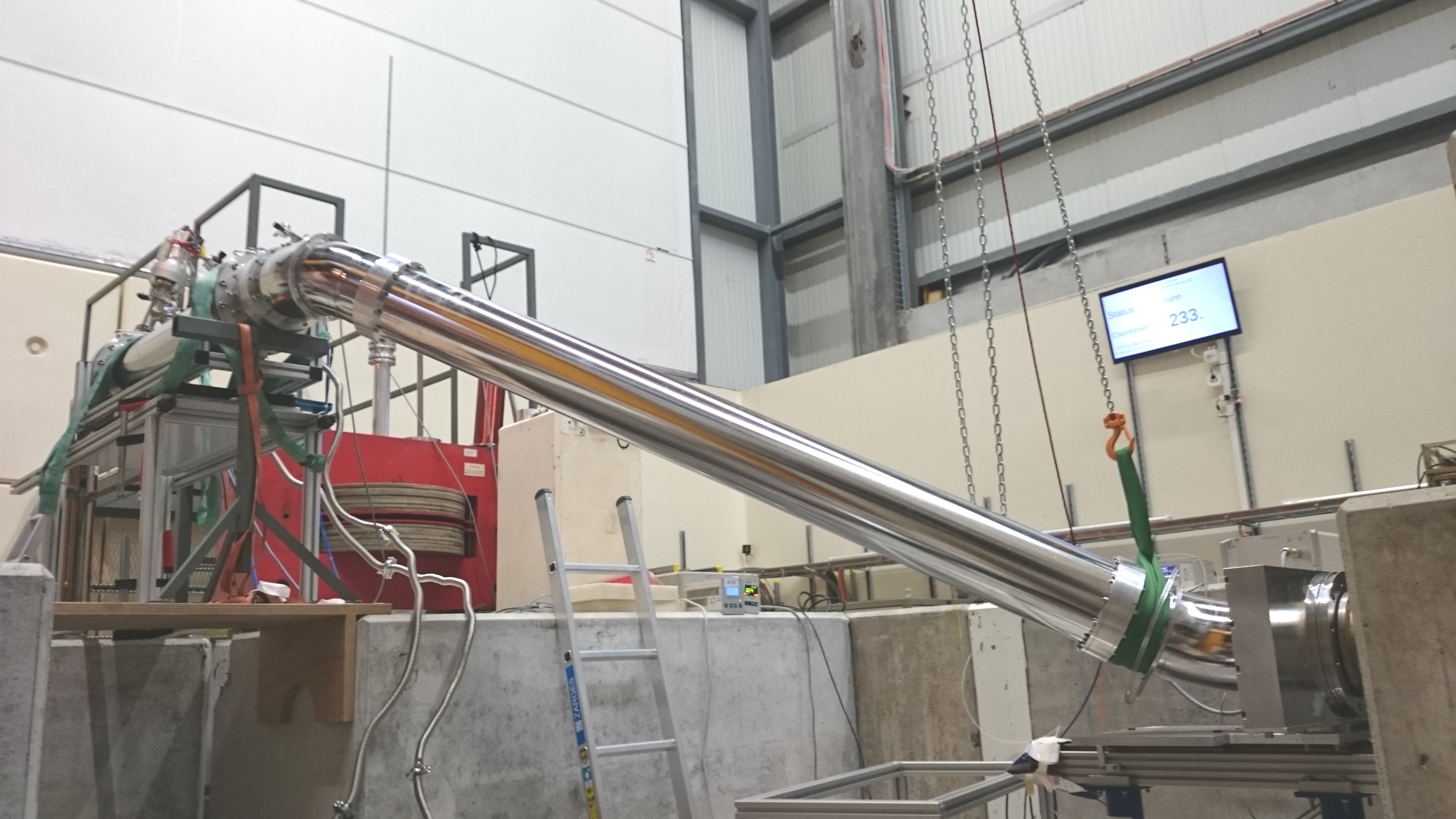}}
\caption[Storage bottle setup at West-1]{Setup with the standard storage bottle elevated by \SI{1200}{\milli\meter} above beamport West-1. }
\label{DensityHighLow}
\end{center}
\end{figure*}

\begin{figure*}[htb]
\begin{center}
\resizebox{0.70\textwidth}{!}{\includegraphics{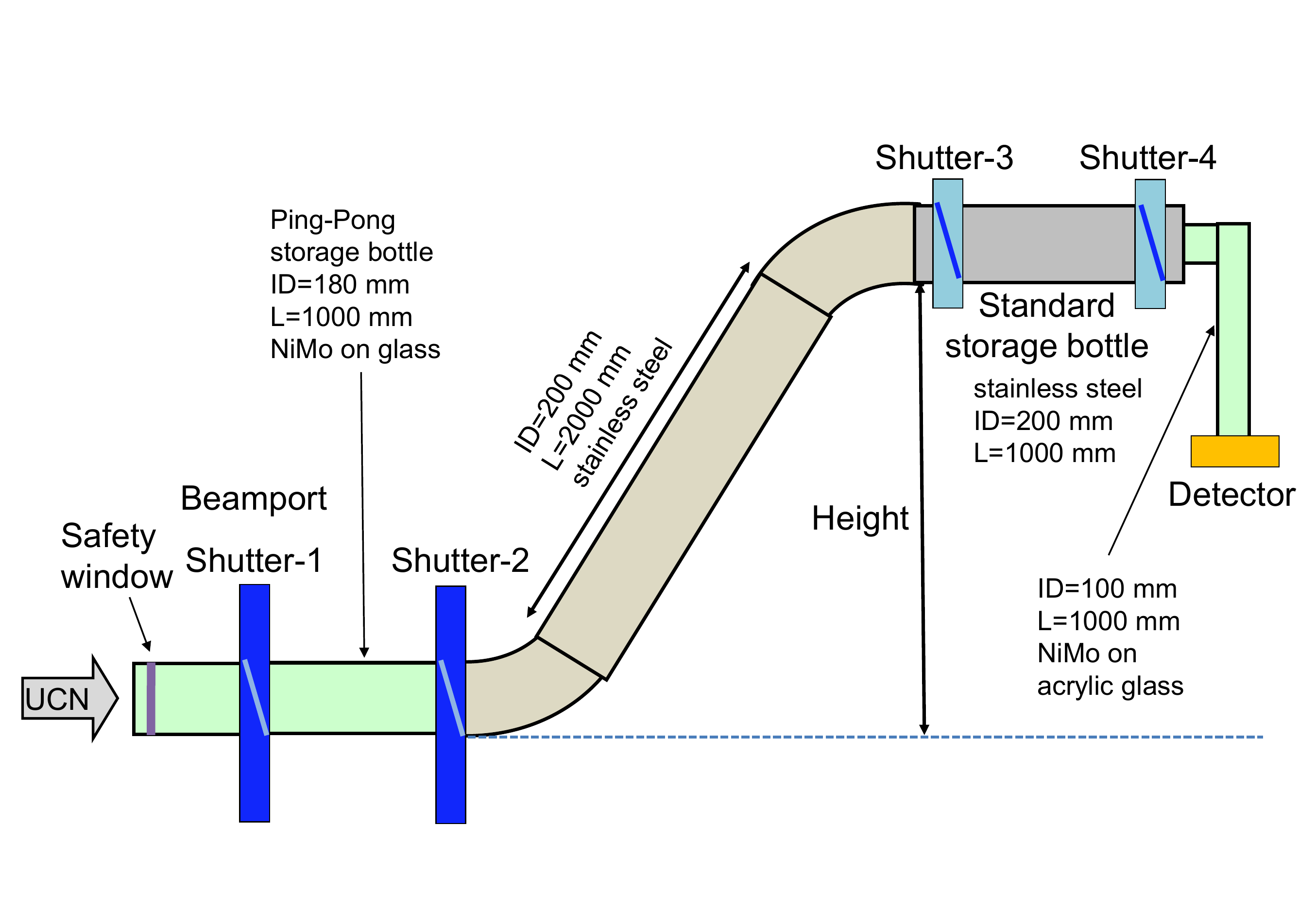}}
\caption[Schematic of the bottle setup at West-1]{The setup for the storage measurement was attached to beamport West-1 replacing the
short guide and detector in Fig.~\ref{fig:ping-pong-setup} after shutter 2.
There, a crank-shaped guide and the standard storage bottle including shutter 3, shutter 4 and the detector were 
mounted.}
\label{Setup_Storage_Bottle}
\end{center}
\end{figure*}

The \SI{\sim31.4}{\liter} standard storage bottle, made from a cylindrical stainless steel tube with inside diameter of 200\,mm and a length of 1000\,mm, could be be mounted at various positions and heights, as shown in Figs.~\ref{DensityHighLow} and~\ref{Setup_Storage_Bottle}.
A crank-shaped UCN guide made from electropolished stainless steel (type 316L) tubes was used to connect the standard storage bottle to the beamport. It consisted of two \SI{45}{\degree} bends and a \SI{2}{\meter} long straight tube. 
By rotation of the second \SI{45}{\degree} bend around the first \SI{45}{\degree}  bend, heights in the range -50\,mm to +1700\,mm with respect to the beamport could be selected with only one set of neutron guides.
The inner diameter of all parts was \SI{200}{\milli\meter} with \SI{2}{\milli\meter} wall thickness.
They were specified as DIN 11865/11866  hygiene class H3, but in addition were electropolished to reduce the already low surface roughness.

A sketch of the setup is shown in Fig.~\ref{Setup_Storage_Bottle} along with the additional shutters 3 and 4. The small Cascade detector was mounted on an L-shaped tube with \SI{100}{\milli\meter} inside diameter, made of acrylic glass,  behind shutter 4. This NiMo-coated UCN guide allowed a 1m fall of UCNs in order to increase their energy accordingly and hence increase transmission through the aluminum entrance window of the detector.

\subsubsection{Measurements}

The measurements were performed over a period of several days and thus the UCN source performance changed during this time, e.g., due to changing proton current or sD2 surface degradation~\cite{Anghel2018}. 
%However, we took special care that the sD$_2$ conditions were very similar by preparation of the sD$_2$ via conditioning~\cite{Anghel2018}. 
In order to compensate for these changes, the  measurements were scaled such that the counts measured simultaneously at beamport West-2 would match the value of \SI{2e6}{} UCNs/pulse, an average value measured at beamport West-2 during this period. In addition, in Ref.~\cite{Anghel2018} it was found and explained that the ratio of count rates between beamports  West-1 and West-2 is constant for a certain integrated beam current on target after conditioning and then decreases linearly over time as visible in Fig. 8 of Ref.~\cite{Anghel2018}. 
This behavior was parametrized and used as correction for the West-2 normalization, however with an assumed large  absolute error of up to 0.05 on the correction factor. 
In most of the cases, measurements were repeated three times at a specific height. 
Proton beam pulses of 5.4\,s length every 300\,s, the longest ones allowed at the time of these measurements, were used with a proton beam current of \SI{2.2}{\milli\ampere}.
In order to minimize the leakage of UCNs from the source into the standard storage bottle during storage, shutters 2 and 3 were closed simultaneously. A \SI{1}{\meter} guide connecting the beamport and shutter 2, namely the storage bottle from the ping-pong measurement described in Sec. 2.1, were part of the beamline at the time of the measurements.
The leakage from the volume enclosed between shutters 2 and shutter 3 was measured separately at every height. In this leakage measurement shutter 3 was closed during the ``filling time'' and shutter 4 was open all the time.

\subsubsection{Results and discussion}

\begin{figure*}[htb]
\begin{center}
\resizebox{0.50\textwidth}{!}{\includegraphics{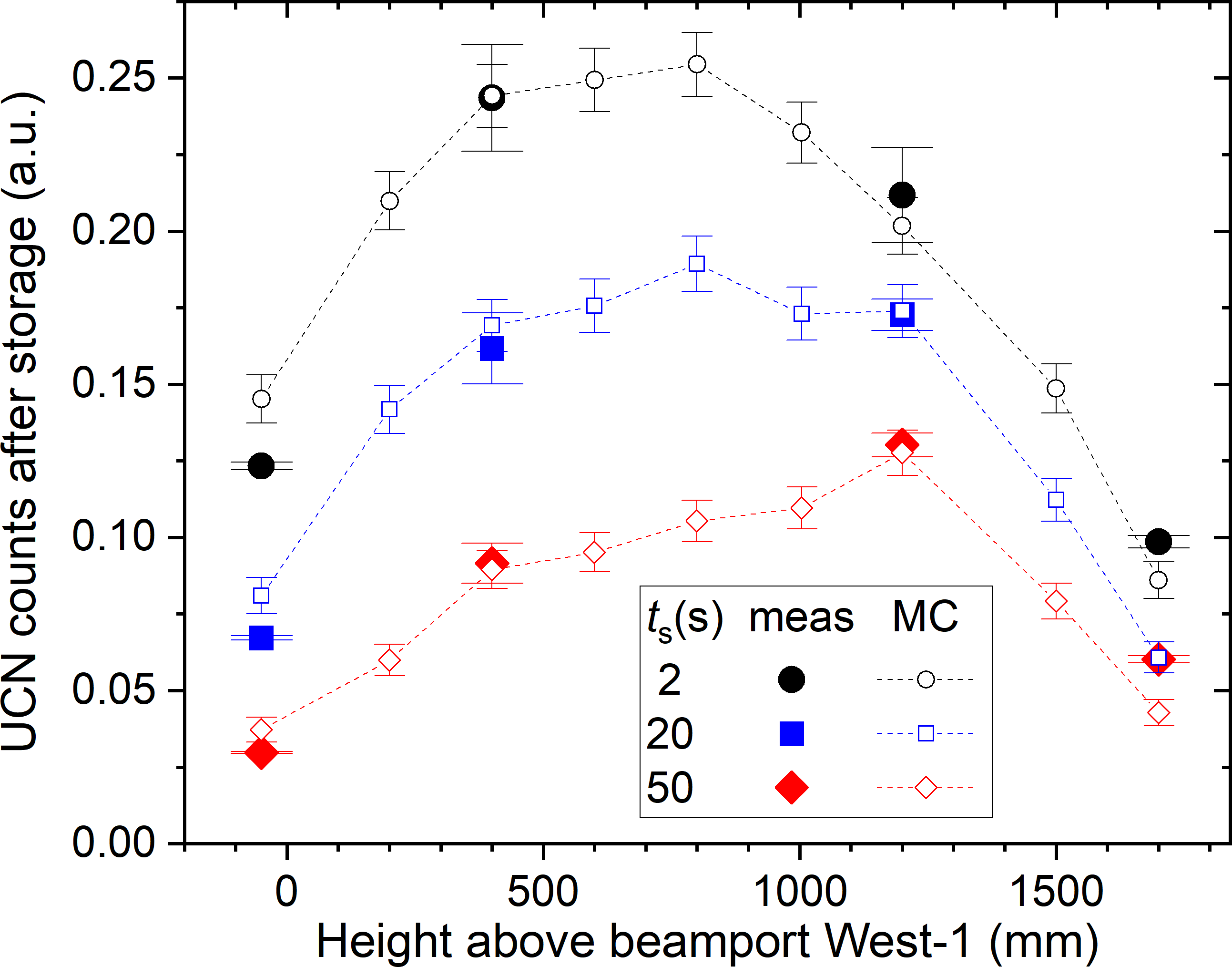}}
\caption[UCN counts at various heights]{UCN counts after $t_s =$ 2, 20 and 50\,s  of storage in the standard storage bottle as measured at various heights above the beamport. The full symbols represent the measured data and the open symbols the simulation (MC), see Sec.~\ref{Sec:simulation}. The counts were normalized as described in the text. }
\label{SteelBottleCountsVsHeight-2s-20s-50s}
\end{center}
\end{figure*}

\begin{figure*}[htb]
\begin{center}
\resizebox{0.50\textwidth}{!}{\includegraphics{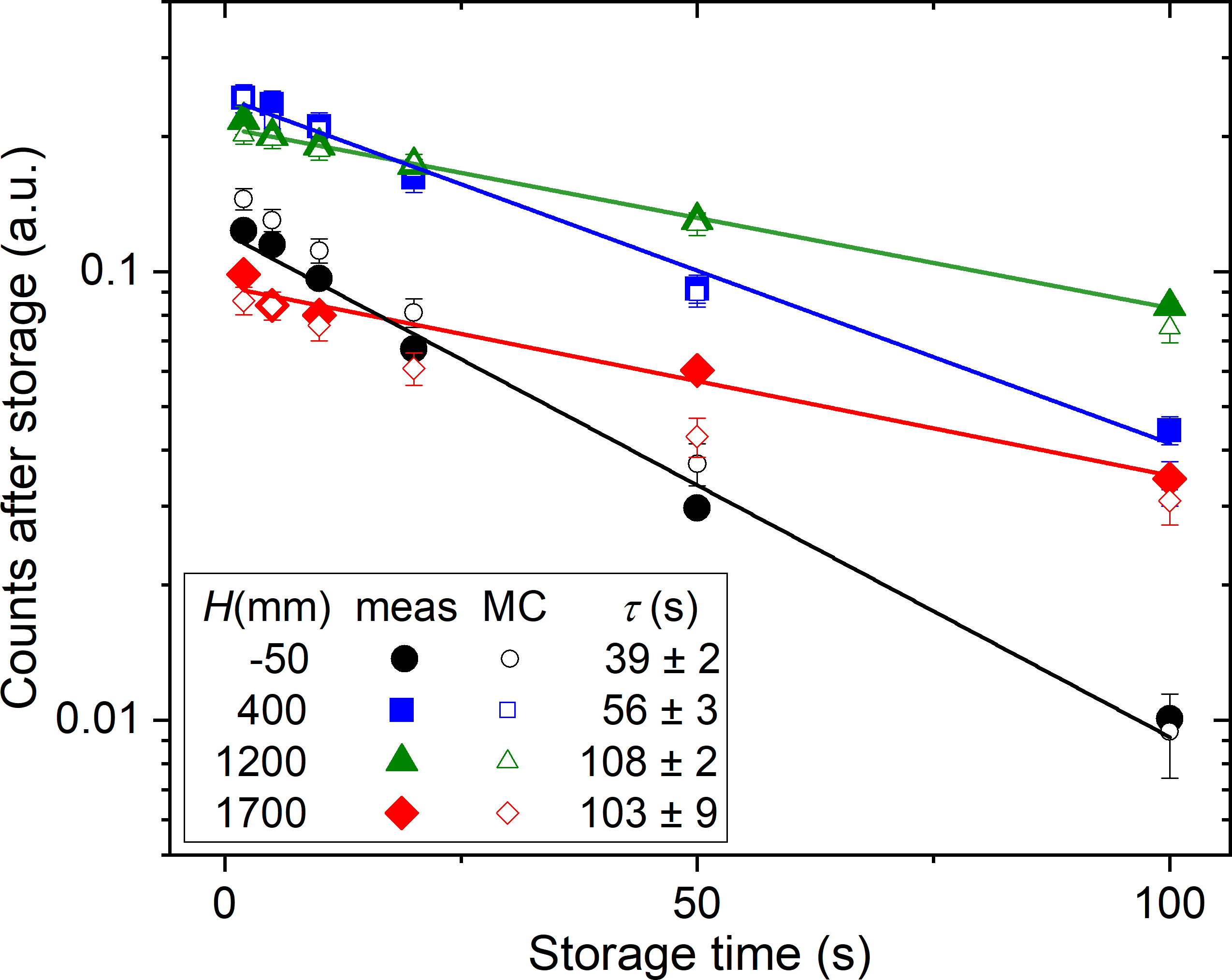}}
\caption[Storage curves at various heights]{Corrected UCN counts (full symbols: measurement; open symbols: MC simulation) at various heights ($H$) above beamport West-1 versus storage time. The lines are one-exponential fits to the measurements with indicated storage time constants, $\tau$.}
\label{SteelBottleStorageOneCommonScalingMCmeas}
\end{center}
\end{figure*}

The height dependence of the UCN density measured after storage times of \SI{2}{\second}, \SI{20}{\second}, \SI{50}{\second} is plotted in Fig.~\ref{SteelBottleCountsVsHeight-2s-20s-50s}. The corresponding storage curves can be seen in Fig.~\ref{SteelBottleStorageOneCommonScalingMCmeas}. The detected UCN counts after \SI{2}{\second} storage and the corresponding corrections are compiled in 
Table.~\ref{HeightCountsTable}. 
%Multiple measurements at the same height demonstrate a sufficient reproducibility obtained with the normalization method using the West-2 counts.  
As mentioned above, the leakage counts were measured at every height and subtracted from the counts of the regular measurement and then normalized and corrected with the West-2 counts obtained in the same proton beam pulse.

\begin{table}[b]
\centerline{
    \begin{tabular}{c|c|c|c|c|c}
%    Height (mm)  & UCN counts after storage & UCN counts at West-2 & density (UCN/cm$^3$) \\\hline
%   \SI{1200\pm10}{} & \SI{321876}{} & \SI{1554499}{} & \SI{13.1}{} \\
%   \SI{1200\pm10}{}   & \SI{244425}{} & \SI{1224614}{} & \SI{13.9}{} \\
%    \SI{1700\pm10}{}   & \SI{78276}{} & \SI{816374 }{} & \SI{7.9}{} \\
%   \SI{1700\pm10}{}   & \SI{164740}{} & \SI{1678166}{} & \SI{6.3}{} \\
%    \SI{400 \pm10}{}   & \SI{268805}{} & \SI{1171547}{} & \SI{15.5}{} \\
%    \SI{-50 \pm10}{}   & \SI{141879}{} & \SI{1031034}{} & \SI{9.3}{} \\
%    \SI{-50 \pm10}{}   & \SI{287894}{} & \SI{2025039}{} & \SI{7.9}{} \\
Height(mm)  &	Counts West-1	&	Counts West-2	&	Leakage cts.	&	Corr. factor	& 	Density(UCN/cm$^3$)	\\\hline
1700$\pm$10 & 164740$\pm$800	 & 1678166$\pm$8400	 & 4226$\pm$40 	& 0.97$\pm$0.02	& 6.3$\pm$0.2 \\
1200$\pm$10	& 321876$\pm$1600	& 1554499$\pm$7800	& 20993$\pm$110	& 0.94$\pm$0.02	& 13.1$\pm$0.4 \\
1200$\pm$10	& 244425$\pm$1200	& 1224614$\pm$6100	& 16538$\pm$90	& 0.86$\pm$0.05	& 13.9$\pm$1.1 \\
400$\pm$10	& 268805$\pm$1300	& 1171547$\pm$5900	& 27727$\pm$140	& 0.85$\pm$0.05	& 15.5$\pm$1.2 \\
-50$\pm$10	& 287894$\pm$1400	& 2025039$\pm$10100	& 38213$\pm$200	& 1$\pm$0.02	& 7.9$\pm$0.2 \\
    \end{tabular}
}
    \caption[Stored UCN counts at various heights]{Measured counts per proton pulse at the given height of the standard storage bottle above the beamport West-1. The storage time was 2\,s. The correction factor (Corr.factor) compensated the decrease with time of the UCN source performance after conditioning~\cite{Anghel2018}. 
%The UCN counts after storage were corrected for leakage through shutter 3. 
The UCN density was calculated as described in the text.
}
\label{HeightCountsTable}
\end{table}

The elevated height was measured from the bottom edge of the UCN guide at the exit of shutter 2 to the bottom of the UCN guide at the entrance of the standard storage bottle at shutter 3.
UCN densities between \SI{6}{UCN/\centi\meter^3} and \SI{15}{UCN/\centi\meter^3} were measured after a \SI{2}{\second} storage time. The simulated values peak at a height of around 800\,mm, suggesting that the UCN density reported in Ref.\cite{Bison2017},  measured at 400\,mm was not optimal.
The counts and densities given in Table.~\ref{HeightCountsTable}  were not corrected for detector inefficiency and UCN  transmission of the crank-shaped guide, and not extrapolated to zero storage time.
The values given here were regularly achieved after longer source operation periods.

The storage time constant in any storage bottle depends on the kinetic energy spectrum of the UCNs.  This spectrum in turn depends on the elevation of the volume above the beamport, since the UCNs are slowed down by gravity. This implies lower bounce-rate and thus smaller losses.
The corresponding storage time constants were obtained using the leakage rate of UCNs through shutter 4 during the storage measurements of times up to \SI{100}{\second} and fitting with a single-exponential function.

%a function of the form

% \begin{equation}
% N(t)\propto R(t) = A_1\times e^{-t/\tau_1} + A_2\times e^{-t/\tau_2},
 %\end{equation}
% where $N(t)$ is the total count of UCN inside the storage bottle, $R(t)$  is the
% rate of UCNs leaking into the detector, $A_1$ and $A_2$ are population
% weights, and $\tau_1$ and $\tau_2$  the time constants of the exponential decay.

% In case of the \SI{1200}{\milli\meter} elevation, the double exponential fit didn't converge, and thus a single exponential decay was used. 
%The reduced $\chi^2$-s of all fits varied between \SI{1.02}{} and \SI{1.16}.
% The resulting storage time constants are listed in Table~\ref{HeightStorageTimeTable}.

% \begin{table}[b]
% \centerline{
%     \begin{tabular}{c|c|c|c|c|c}
%     Height (mm) & $\tau_1$ (s) & $A_1$ & $\tau_2$ (s) & $A_2$ & $\chi^2/% n_\text{df}$ \\\hline
%    \SI{-50}{} & \SI{14.9\pm 2.0}{}& \SI{125\pm21}{} &  \SI{41.5 \pm 2.1}{} & \SI{200\pm22}{} & \SI{1.14}{} \\
%    \SI{400}{}   & \SI{23.8\pm 4.5}{}& \SI{147\pm43}{} &  \SI{70.9 \pm 10.0}{}& \SI{190\pm45}{} & \SI{1.02}{}  \\
%    \SI{1200}{}  & ---               & --- &  \SI{101.9\pm 1.1}{} & \SI{225\pm1}{} & \SI{1.03}{} \\
 %   \SI{1700}{}  & ---               & --- &  \SI{92.2 \pm 1.2}{} & \SI{135\pm1}{} & \SI{1.16}{} \\
%     \end{tabular}
% }
%     \caption[Storage time constants at various heights]{Storage time constants, % from a double or single exponential fit to the data, corresponding to various elevations of the storage bottle above the West-1
%     UCN beamport.}
%     \label{HeightStorageTimeTable}
% \end{table}

Fig.~\ref{SteelBottleStorageOneCommonScalingMCmeas} summarizes the storage curves obtained at various heights, 
% including reproducibility measurements and 
applying the described normalization method. The legend contains the storage time constants, $\tau$, from an exponential fit to the measurement. We use only a single exponential fit to the data in order to demonstrate the expected behavior that, at a higher position of the standard storage bottle, the average UCN energy is smaller. The storage time constants at 1200 and 1700 mm heights agree within errors, indicating no relevant change of the spectrum.

The UCN transmission of the crank-shaped UCN guide was measured in storage mode by 
comparing the horizontal storage measurement to a storage measurement 
without the crank-shaped guide. At the height of the the beamport, a transmission of  0.72$\pm$0.02 was found for different storage times between 2 and \SI{20}{\second}.
%The transmission became higher for much longer storage times. This is due to the fact that longer storage times result in a softer UCN spectrum corresponding to a lower bounce rate in the S-shaped guide during the filling procedure.

%After the measurements at various heights were finished, the storage bottle setup was mounted at a height of about \SI{500}{\milli\meter} above the West-1 beam port using the identicel \SI{45}{\degree} stainless steel bends but only a \SI{1}{\meter} long straight stainless steel guide. This measurement is described in detail in Sec.~\ref{densityComparisonMeasurementPSI} and yielded a UCN density of \SI{22.31\pm0.71}{UCN\per\centi\meter^{3}} after \SI{2}{\second} of storage.

%\clearpage

%%%%%%%%%%%%%%%%%%%%%%%%%%%%%%%%%%%
\section{Further analysis of the experimental results by Monte Carlo simulations}
\label{Sec:simulation}

The MCUCN code used in this work was developed at PSI. The code and its physical parameters are described in detail in~\cite{Zsigmond2018}. The simulation model of the UCN source volume and beam guides was benchmarked with a first set of test experiments as reported in~\cite{Bison2020}. 
%There, both the loss per bounce parameter and diffuse reflection probability per bounce for the coating and surface of the installed UCN guides were constrained from measured emptying time-of-flight spectra, while the optical potential parameters were taken from previous publications. In order to only concentrate on the UCN energy range, which will be later interesting in fundamental physics experiments, we were examining arrival times around several tens of seconds, which correspond to typical filling times. 
%In the follow-up experiments, described in this paper, we used stored UCNs in order to again focus onto the energy range relevant for future UCN applications i.e. also to match the effective model parameters to this range.
For a better overview, we give here the list of experiments that were considered in the new benchmark of the model: (i) ping-pong (Sec.~\ref{pingpongexperiment}),  (ii) time spectra from direct emptying of the UCN source vessel from reference~\cite{Bison2020}, (iii) UCN storage with NiMo-coated bottle reported in Table F.8 of~\cite{Goeltl2012}, (iv) standard storage bottle as a function of height (Sec.~\ref{SteelBottleHeights}), (v) dedicated wall loss measurements~\cite{Bondar2017}, and (vi) transmission through AlMg3 foils for the safety windows as reported in~\cite{Atchison2009} and confirmed in  measurements at PSI, see Fig.~28 in~\cite{Bison2020}. 

Experiments (i) - (iv) were used to determine the wall loss coefficient and the probability of Lambert diffuse reflections~\cite{Golub1991} in the beam guides as free parameters.  
Results reported in (v) were used as  a complementary constraint on the wall loss coefficient. The transmission loss in AlMg3 from experiment (vi) was used as an input parameter.

We need two independent model sections for the simulation of the two measurement setups described in the previous chapters. Both are added  after the beamport in separate simulations, attaching them to  the first part of the MCUCN model that includes the entire system of UCN source and its UCN guides. On the one hand, the  coating parameters of the guides up to the beamports $-$ constrained from the ping-pong experiment $-$  influence the energy spectrum of the UCNs which can be stored in the second experiment, with the standard storage bottle placed at various heights. On the other hand, this energy spectrum will determine the transmission of UCNs which is detected in the ping-pong experiment.
Therefore, in the simulations we performed several iterations in scanning the parameter space  aiming to minimize the deviation between measured and simulated UCN counts.

\subsection{Simulation model of the ping-pong measurement}
\label{Sec:simulation-ping-pong}

The MC simulation of the ping-pong measurement followed the same steps as taken in the experiment: 
8.6\,s filling of the external storage volume, 
130\,s of UCN storage, 
cleaning the source volume via the main flapper valves, and
counting of UCNs released after storage towards the detector at beamport South. 
In the simulation model the reference counts were simulated by opening shutter 2 towards the big Cascade detector at beamport West-1 and using this to fit the model parameters to best match the corrected fractions from the measurement listed in Table~\ref{tab:PinpongResTable}. 

The time-dependent fraction of the opened area of the two UCN shutters of the external storage vessel~\cite{Goeltl2012,Bison2016} was implemented in the simulation model. The simulated opening function reproduced the measured one well. 
Furthermore, the field of the polarizing magnet was simplified to a rectangular potential profile along the beam axis located at the position of the safety window, as justified in~\cite{Bison2020}.

%As in our previous study~\cite{Bison2020}, we based on earlier simulations considering the full 3D map of the magnetic field,  in which we tested whether the radial field gradient in the 5\,T  polarizer magnet at the South beamport would have a non-negligible effect on the UCN transmission through a guide placed in its bore. We concluded that the radial field gradient in the magnet has a negligible effect on  the UCN transmission, comparable to the simulation errors. Therefore, the field of the magnet in the final MCUCN model was simplified to a rectangular potential profile along the beam axis located at the position of the safety window. This strongly reduced the otherwise excessive computation time, and enabled to scan the parameter space within several weeks.  Hence we obtained acceptable simulation statistics which can be compared to the one of the measurements.

The parameters of the source storage vessel were set to the value obtained in~\cite{Bison2020}. Changing the fraction of Lambert diffuse reflections in the source storage vessel in a range 10\% - 50\% had no discernible influence on the following  results.
The simulated transmission efficiency of the UCNs propagated from beamport West-1 to beamport South  shows a strong dependency on two parameters: 
the loss-per-bounce parameter $\eta_\text{source guides}$ and the Lambert diffuse fraction of reflection in the two UCN guides, $p_\text{diff,source guides}$. 

\begin{figure*}[ht]
\begin{center}
\resizebox{0.50\textwidth}{!}{\includegraphics{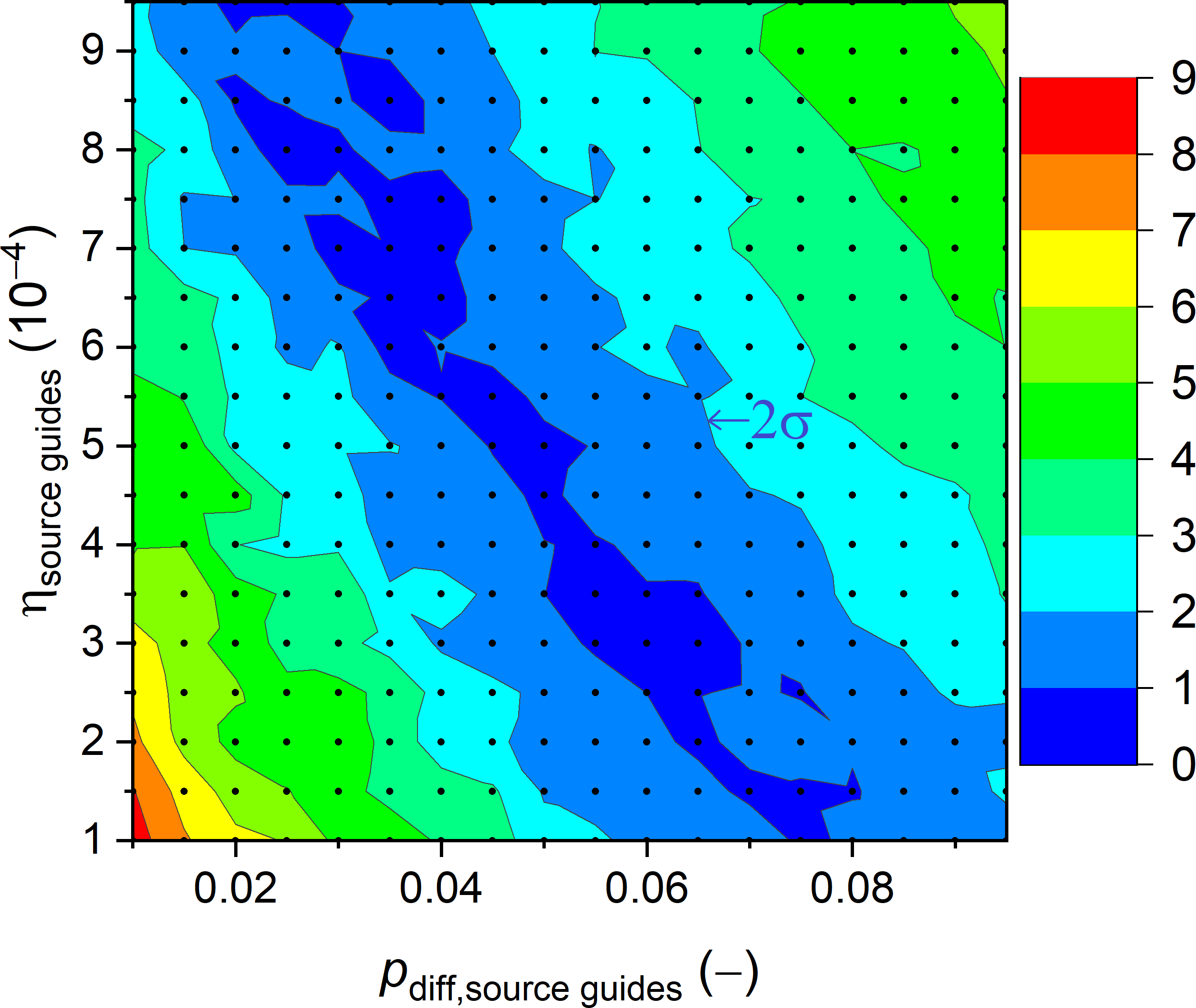}}
\caption{MC scans of the relative deviation from experiment in the $\eta_\text{source guides}-p_\text{diff,source guides}$ parameter plane of the beam guides combining the cases polarizer  magnet off and on. The color scale shows the deviation in 1$\sigma$ unit, the 2$\sigma$ second darkest border is indicated. 
%The horizontal lines at $3.0\pm0.5\times$10$^{-4}$ indicate the value and its uncertainty obtained from~\cite{Bondar2017}, 2$\times$10$^{-4}$ but adding 1$\times$10$^{-4}$, originating from the gaps along the guides, $\approx$0.1 mm/meter.
} 
\label{fig:PingPong_2D-FractionSCM0.0Tand5.0T}
\end{center}
\end{figure*}

We performed MC scans in the 2D parameter space, see Fig.\,\ref{fig:PingPong_2D-FractionSCM0.0Tand5.0T}. 
The runtime of one configuration lasted several hours, thus we calculated a grid of parameters, and interpolated with triangular interpolation. 
The goodness of the MC fit was estimated by calculating the quadratic deviation from the measured counts in detector South and detector West-1. This deviation was expressed  in units of the square root of the quadratic sum of the errors from the simulation and measurement (1$\sigma$). The result is shown in Fig.\,\ref{fig:PingPong_2D-FractionSCM0.0Tand5.0T} combining  the cases with polarizer  magnet off and on quadratically. The color scale shows the deviation in units of 1$\sigma$.

We identify in the 2D plot a linearly-shaped valley displaying a linear anti-correlation between the loss parameter and the fraction of diffuse reflections. 
%
%A high $\eta_\text{source guides}$ and low $p_\text{diff,source guides}$ is equivalent to a low $\eta_\text{source guides}$ and high $p_\text{diff,source guides}$. 
%
%This 2D parameter constraint can be overlapped with the measurement of the loss parameter for NiMo coatings from Ref.~\cite{Bondar2017}, 2$\times$10$^{-4}$, but adding 1$\times$10$^{-4}$ corresponding to the gaps along the guides, of about 0.1 mm/meter, and estimating a common uncertainty of 0.5$\times$10$^{-4}$.
%

\begin{figure*}[ht]
\begin{center}
\resizebox{0.50\textwidth}{!}{\includegraphics{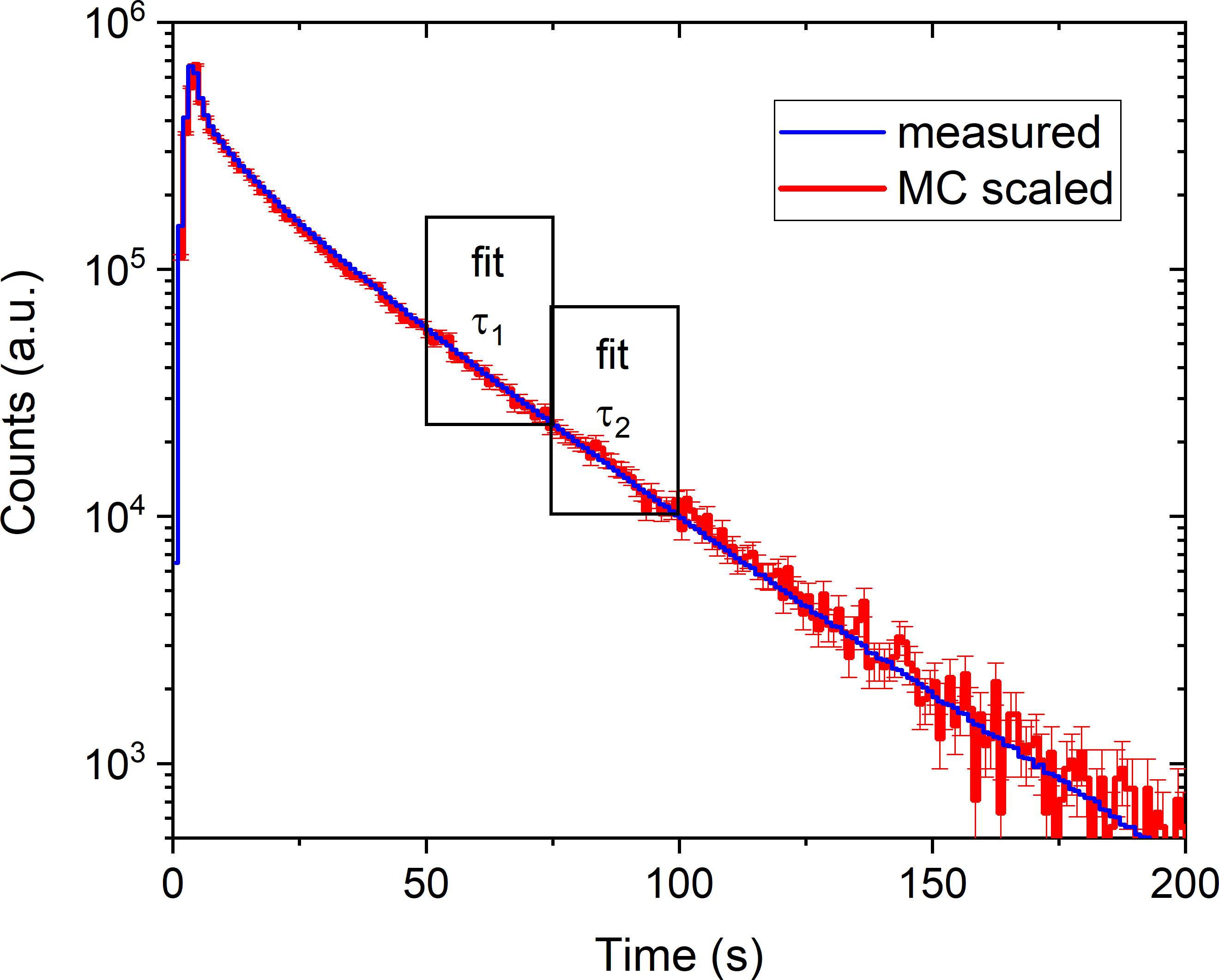}}
\caption{Measured and simulated time profiles of UCN counts observed with the big Cascade detector at beamport South.  In the simulation $\alpha=2.7$ for the initial spectrum. The rectangles indicate the fit intervals in which the time constants, used in the comparison, are determined.} 
\label{fig:PINGPONG2020_EmptySouthTauPow2.7-TOFplot}
\end{center}
\end{figure*}

\begin{figure*}[ht]
\begin{center}
\resizebox{0.50\textwidth}{!}{\includegraphics{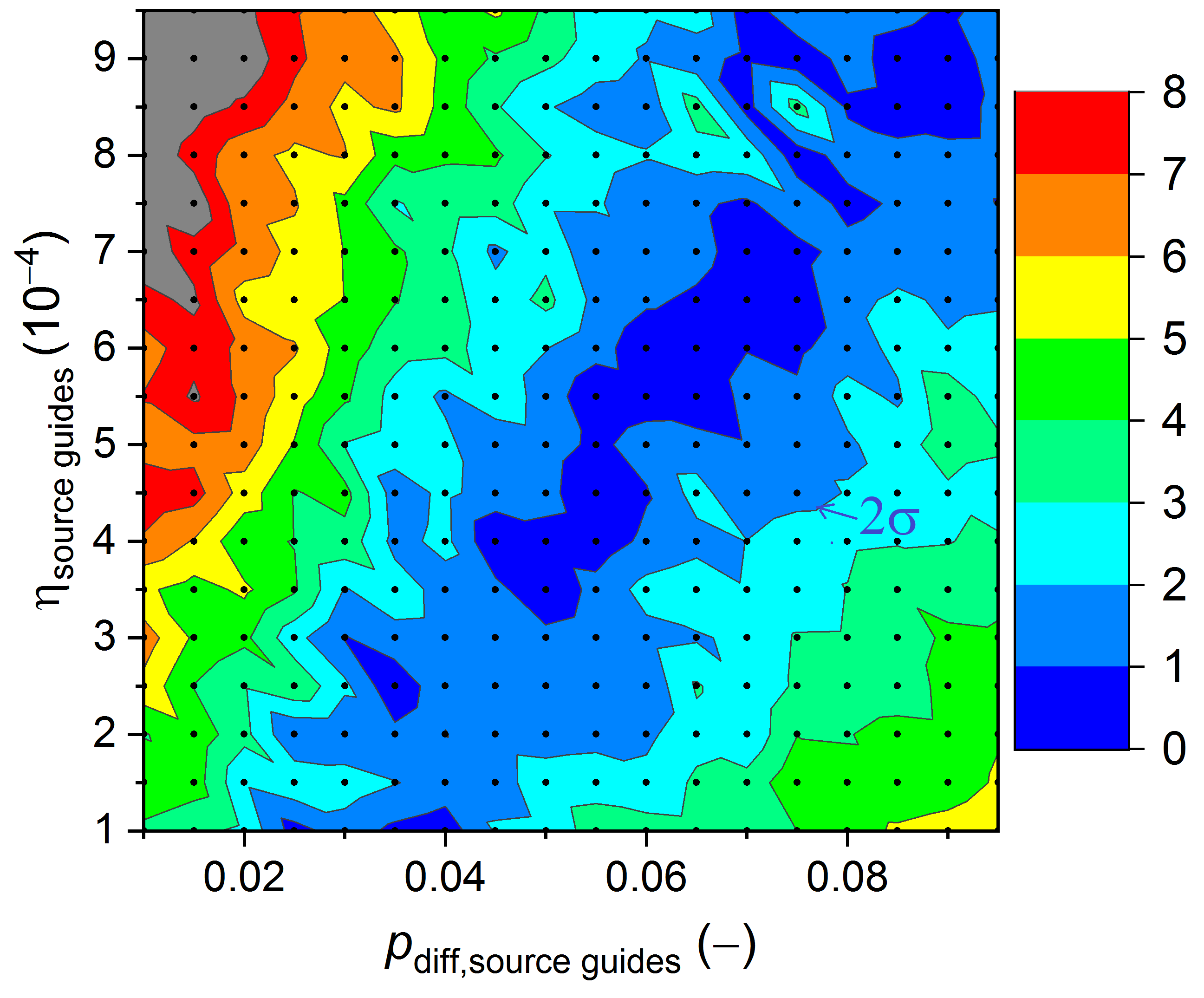}}
\caption{MC-scan obtained  from fitting the time profile of UCN counts observed with the detector at beamport South (Fig.~\ref{fig:PINGPONG2020_EmptySouthTauPow2.7-TOFplot}). $\alpha=2.7$ is set for the initial spectrum. The color scale shows the deviations in 1$\sigma$ unit, the 2$\sigma$ second darkest border is indicated. 
%The horizontal lines at $3.0\pm0.5\times$10$^{-4}$ indicate the value and its uncertainty obtained from~\cite{Bondar2017}, 2$\times$10$^{-4}$, but adding 1$\times$10$^{-4}$, originating from the gaps along the guides, $\approx$0.1 mm/meter.
} 
\label{fig:DirectEmptyingSouth_2D-Tau}
\end{center}
\end{figure*}

A second constraint was obtained from a very similar parameter scan in which we fitted the emptying time-profile measured in detector South, see Figs.\,\ref{fig:PINGPONG2020_EmptySouthTauPow2.7-TOFplot} and \ref{fig:DirectEmptyingSouth_2D-Tau}. This was done similarly as in~\cite{Bison2020} using two time constants, $\tau_1$ and $\tau_2$, separately for two consecutive time ranges, see rectangles in Fig.\,\ref{fig:PINGPONG2020_EmptySouthTauPow2.7-TOFplot}, where we expect only UCNs, i.e. neutron energies smaller than 250 neV (see Fig.\,\ref{fig:spectrum_cleaning_after_12s}) but still have sufficient statistics. In this work we generated an initial energy spectrum from the sD$_2$ moderator by using a free parameter, $\alpha$ as exponent: $P(E)dE\propto E^{\alpha} dE$, in contrast to a linear spectrum ($\alpha=1$) as assumed in Ref.~\cite{Bison2020}. This can be motivated by a smaller extraction efficiency for the slowest UCNs due to for example back-reflections expected at crack walls in a polycrystalline sD$_2$.

\begin{figure*}[ht]
\begin{center}
\resizebox{0.50\textwidth}{!}{\includegraphics{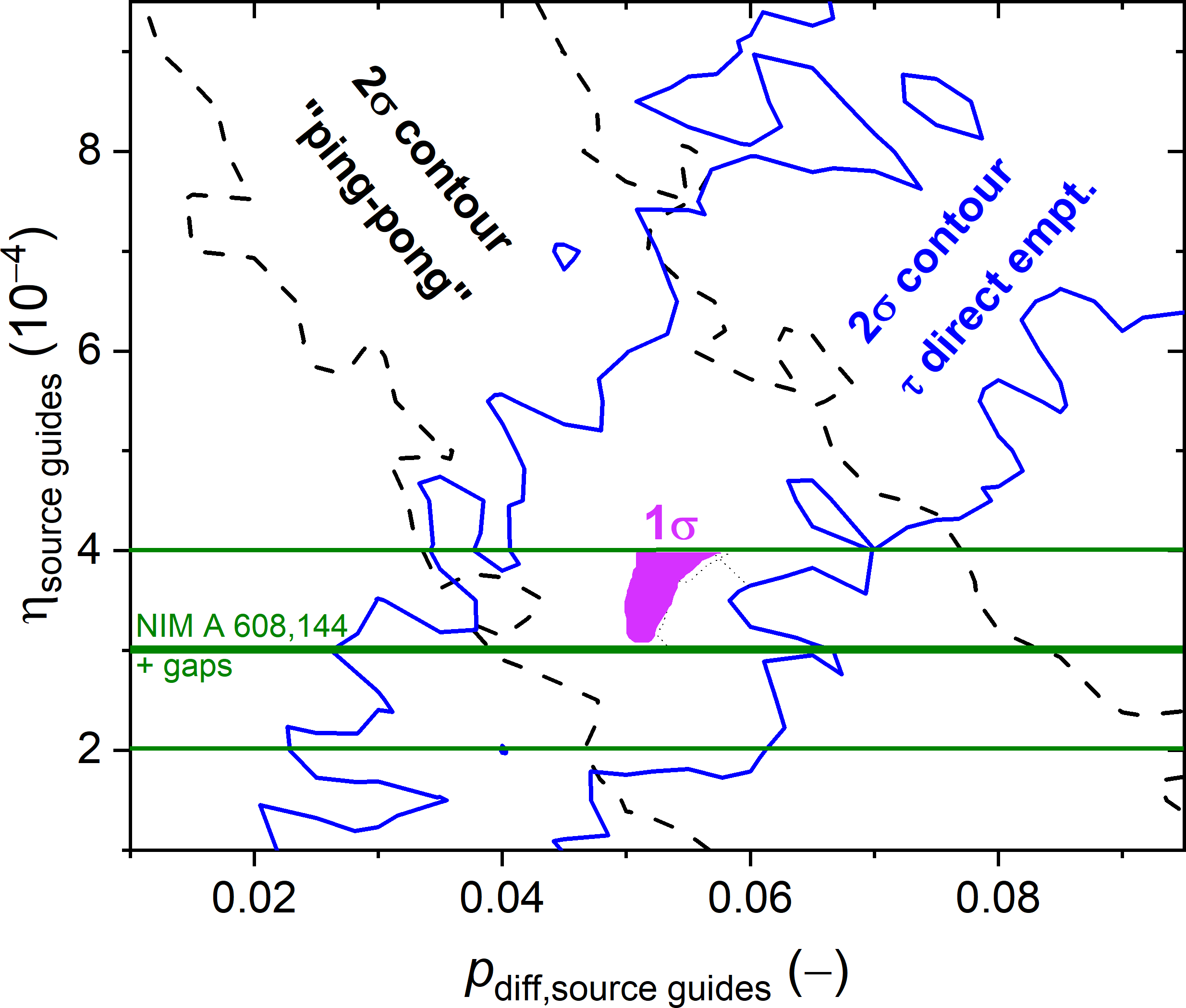}}
\caption{Combination of the 2$\sigma$ regions separate from (i) the ping-pong (Fig.\,\ref{fig:PingPong_2D-FractionSCM0.0Tand5.0T}) and (ii) the direct emptying (Fig.\,\ref{fig:DirectEmptyingSouth_2D-Tau}) represented by the contours of the aforementioned constraints. The horizontal lines at $(3\pm1)\times10^{-4}$ indicate the value and its uncertainty obtained from~\cite{Bondar2017} but adding a contribution from possible gaps along the guides (see text). The violet spot indicates the region where all three constraints overlap within their 1$\sigma$.} 
\label{fig:Combine-2sigma-contours-PingPong-and-Tau_direct_emptying}
\end{center}
\end{figure*}

The combined results from the previous constraints are shown in Fig.\,\ref{fig:Combine-2sigma-contours-PingPong-and-Tau_direct_emptying}. 
This resulting parameter constraint can be overlapped with a third one from the measurement of the loss parameter for NiMo (85/15 weight percent) coatings from Ref.~\cite{Bondar2017}, $(2.0\pm0.4)\times10^{-4}$. There we added $(1\pm1)\times10^{-4}$ from a rough estimation of perfectly absorbing gaps at the guide flanges, which cannot be excluded. The estimation of 100\% absorbing gaps and its uncertainty is based on mechanical measurements after guide installation at the UCN source, consistent with analysis results in Ref.~\cite{Goeltl2012}.

The violet area in Fig.\,\ref{fig:Combine-2sigma-contours-PingPong-and-Tau_direct_emptying} indicates the deviations around $p_\text{diff,source guides} = 0.05$, where all three constraints overlap within 1$\sigma$. This relatively high value can be explained by the non-ideal roughness of the stainless steel guide sections consisting of about 20\% and 25\% of the full flight path in the UCN guides South and West-1, respectively. The loss parameter for the NiMo storage bottle, as an independent parameter in the iterations, was estimated to be around $5.5\times10^{-4}$, effectively including gaps from the shutters.

In a next set of simulations, the count fractions in detector South as a function of the magnetic field in the polarizer were compared to the measurements. The results were compiled in Table~\ref{tab:PinpongResTable} and plotted in  Fig.~\ref{fig:PingPongFractionSouthWestvsSCfieldPow2.7}. The simulated values were obtained by using the final fit parameters: a  fraction of diffuse reflections of 0.05, a loss parameter 3$\times$10$^{-4}$ and an exponent $\alpha$=2.7 in the initial energy spectrum of UCNs exiting the sD$_2$, obtained in the parameter fit discussed in  Sec.~\ref{Sec:simulation-storage}. 
The simulation fits the measured data sufficiently well.
%, except at 1.5\,T with 3$\sigma$ discrepancy. However, using global coating parameters for a number of individual surfaces also around the magnet position, may be a cause of the different profile of the simulated curve, most conspicuous at 1.5\,T, and will be further investigated.

\begin{figure*}[ht]
\begin{center}
\resizebox{0.50\textwidth}{!}{\includegraphics{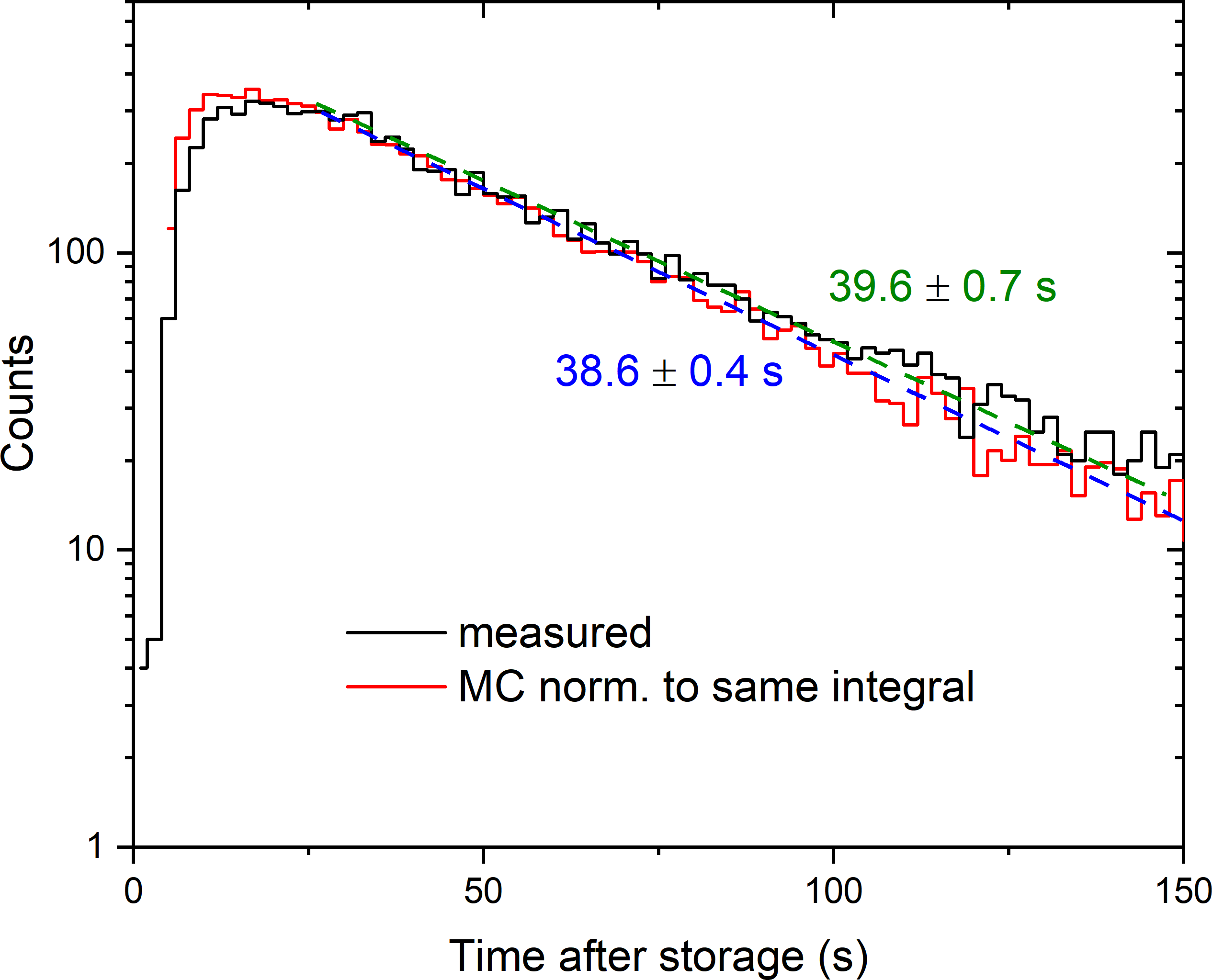}}
\caption{Observed and simulated UCN time spectra in the ping-pong measurement with the big Cascade detector at beamport South. The green and blue labels indicate the single-exponential time constants of the measured and simulated data, respectively.} 
\label{fig:PingPong_TimeProfilesMCandMeasured}
\end{center}
\end{figure*}

As a cross-check of the ping-pong simulation results, independent of the MC fitting procedure, we compared the measured and simulated time-profiles in the ping-pong experiment. For this we used  the final fit parameters. 
The simulated peak in the detector at beamport South, shown in Fig.\,\ref{fig:PingPong_TimeProfilesMCandMeasured}, is located about 2\,s prior to the measured peak position.
This is probably due to the fact that on a short time-scale below 25\,s, UCNs experience only a  small  number of diffuse reflections. Thus the details of the diffuse reflection model are important. In this short time range, utilizing the Lambert model that implies ``memory loss'' for the trajectories after a single diffuse reflection, instead of e.g. a smearing of the specular angle of reflection, seems to be less appropriate. At times considerably larger than the flight time between two diffuse reflections (about 2\,s assuming a fraction of diffuse reflections of 0.05), we obtain a  close match between the emptying time constants from simulation and measurement: $38.6\pm0.4$\,s simulated, and $39.6\pm0.7$\,s measured. 
%For the reasons mentioned just above, the shape of the time spectrum in the West-1 detector, with most of the UCNs arriving at times shorter than 10 s, could not be well reproduced in the simulations.  

%The second crosscheck, the shape of the simulated time spectrum in the West-1 detector when using the small Cascade detector, see Fig.\,\ref{fig:PingPong_TimeProfilesMCandMeasuredWest-1}, needs a new parameter, the losses at  shutter 2 in open position. These losses are favored by reflections on the part reducing the guide diameter that connects shutter 2  to the small Cascade detector. In our simulation approach, we assumed a loss-rate after opening the shutter which persists over the full time of emptying due to a lower optical potential of a small part of shutter 2 seen by UCNs in the open position (e.g. an O-ring).
The transmission of UCNs through AlMg3 foils for the safety windows was measured as reported in~\cite{Atchison2009}. We used the loss factor obtained in~\cite{Atchison2009} as input parameter for the simulations, see Table~\ref{table:SimulationParameters}. However, we also checked whether the simulated UCN transmission through the foil after 130\,s storage time using the loss factor from~\cite{Atchison2009} was consistent with a similar measurement.
In the simulation we obtained a value of 0.60$\pm$0.01, which is 14\% away from the measured transmission.  The latter was 0.70$\pm$0.03, see Fig.~3-13 in Ref.~\cite{Ries2016}. The difference may come on one hand from the limited accuracy of the simulation and on the other hand from the reproducibility in the quality of aluminum foils.

%\begin{figure*}[ht]
%\begin{center}
%
%\resizebox{0.50\textwidth}{!}{\includegraphics{fig-mc/PingPong_TimeProfilesMCandMeasuredWest-1.png}}
%
%\caption{Measured and simulated time spectra for the ping-pong counts in the detector at beamport West-1, assuming a loss due to the reduction of the optical potential of the shutter ring seen by the UCNs with the shutter persisting in open  position.} 
%
%\label{fig:PingPong_TimeProfilesMCandMeasuredWest-1}
%\end{center}
%\end{figure*}
%
%
%******************************************************************
%
\subsection{Simulations of the UCN storage experiment at various heights}
\label{Sec:simulation-storage}

In the MCUCN simulation model of the storage bottle measurement at various heights we reproduced the full storage measurement procedure, including setting the height above the beamport by rotation of the crank-shaped UCN guide, according to details described in the experimental section. 
We used separate parameters for this setup starting at the beamport with the stainless steel guide bend, and the source part upstream of the beamport with parameters determined in the ping-pong measurement.
 
Since we were interested in simulations of relative counts, the vertical detector arm connected by a 90 deg bend to the standard storage bottle was simplified in the model to an ideal UCN counter (without an aluminum foil) just after the diameter-reduction piece. A detailed model of the L-shaped tube would have needed additional uncalibrated parameters e.g. for diffuse reflections and gaps. We assumed a negligible change in the detection efficiency of the measurement at different heights thanks to the about 1\,m vertical acceleration of the UCNs before the detector foil.

The simulated height dependence of the UCN density after \SI{2}{\second}, \SI{20}{\second}, and \SI{50}{\second} of storage is shown in Fig.~\ref{SteelBottleCountsVsHeight-2s-20s-50s}, displaying for the standard storage bottle a flat optimum around 800\,mm at 2\,s storage time.
The simulated storage-time dependence of the UCN counts as a function of the height is shown in Fig.~\ref{SteelBottleStorageOneCommonScalingMCmeas}. 
In both figures all curves have a common factor between measured and simulated counts, obtained from the fit of the simulations to the 2\,s storage measurement shown in Fig.~\ref{SteelBottleCountsVsHeight-2s-20s-50s}.

%The input parameters obtained from previously reported measurements and set fixed in these storage bottle simulations are: (i) the loss parameter in the source guides coated with NiMo set to $3.0\times$10$^{-4}$ is  obtained from 2$\times$10$^{-4}$, our measurements reported in~\cite{Bondar2017} adding 1$\times$10$^{-4}$ representing the gaps of 0.1 mm/meter along the guides (this direct addition is approximately valid for UCN energies around $V_F/2$, see Fig. 2.4 of~\cite{Golub1991}); (ii) the loss cross section in the AlMg3 vacuum separation windows is set to a factor 2.2 more than the loss cross section in pure aluminum corresponding to  earlier  measurements~\cite{Atchison2009}, confirmed in~\cite{Bondar2017}.

\begin{figure*}[ht]
\begin{center}
\resizebox{0.50\textwidth}{!}{\includegraphics{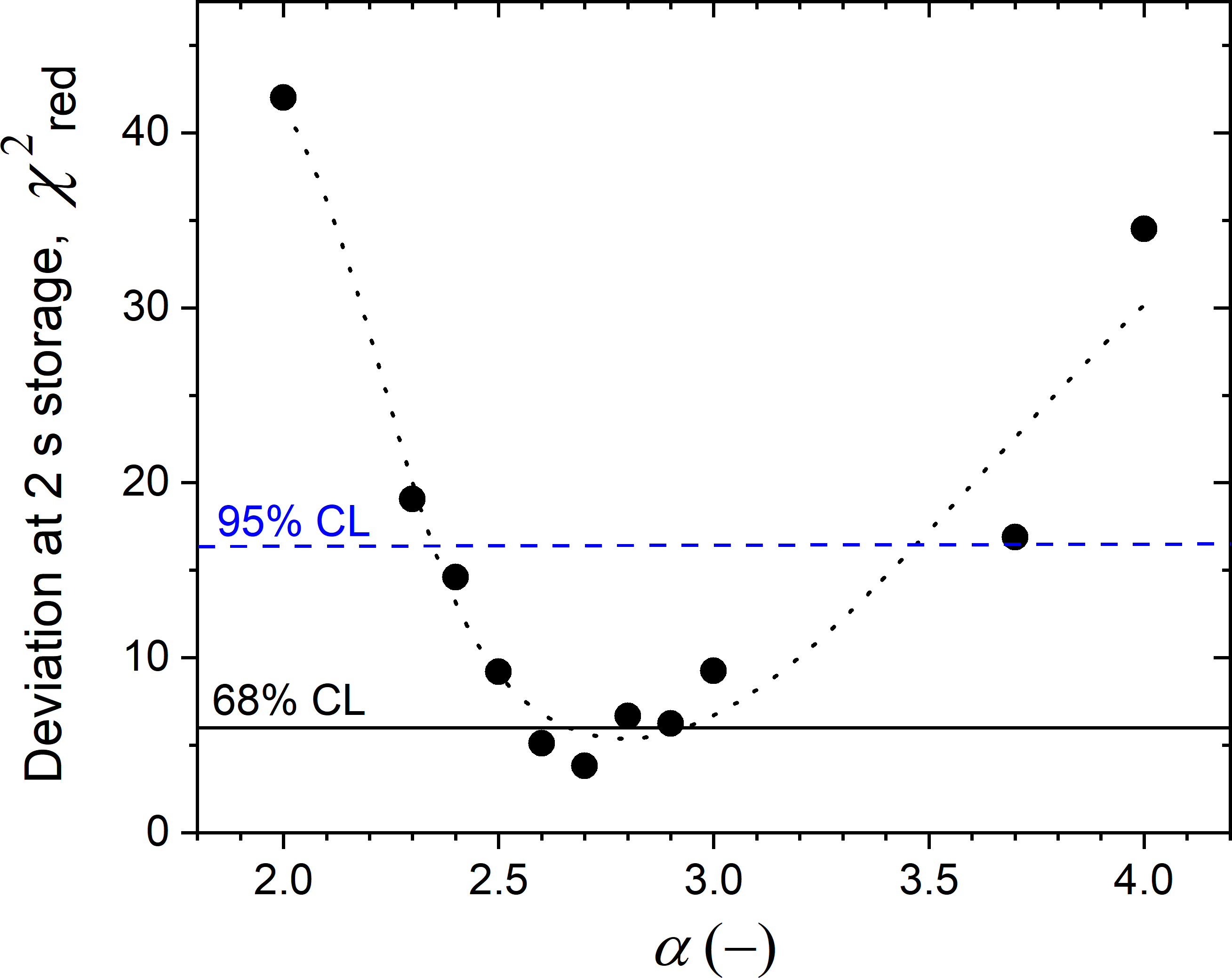}}
\caption{Deviation between the measured and simulated data obtained at 2 s storage time at four different heights (Fig.~\ref{SteelBottleCountsVsHeight-2s-20s-50s}), expressed in $\chi^2_\text{red}$, as a function of the spectral parameter $\alpha$. The dotted line serves as guide to the eye, the horizontal lines indicate the 68\% and 95\% C.L., respectively.} 
\label{fig:SteelBottleAlphaSampling}
\end{center}
\end{figure*}

The fit parameters and their uncertainties for the stainless steel walls and the spectral exponent, $\alpha$, introduced in Sec.~\ref{Sec:simulation-ping-pong}, are obtained from several iterations comparing to the storage measurements at various heights (see summary of all material parameters in Table~\ref{table:SimulationParameters}):
\begin{itemize}
\item a loss parameter for steel , $W/V_\text{F}=(3.4\pm0.5)\times10^{-4}$, along with a gap parameter  (3.0$\pm$0.8)$\times$10$^{-4}$; 
\item an optical potential of 174$\pm$5 neV for steel i.e. about 10\,neV lower than the literature value in Table 9.2 in~\cite{Ignatovich1990}, but within the range for possible different steel alloys and measurement uncertainties;
\item in order to reproduce the measured transmission of the crank-shaped guide, 0.72$\pm$0.02, the parameter for diffuse reflections, $p_\text{diff}$ is set to 0.40 in these guide sections;
\item an exponent $\alpha=2.7\pm0.2$ for the initial energy spectrum of the UCNs  at the surface of the sD$_2$ moderator, obtained via sampling $\alpha$ as shown in Fig.~\ref{fig:SteelBottleAlphaSampling} to simultaneously fit the four data points (four heights) at 2\,s storage time in Fig.~\ref{SteelBottleCountsVsHeight-2s-20s-50s}. The 68\% C.L. was estimated using the Fisher method (F-test) for four data points and one fit parameter~\cite{ROGERS1975}, yielding the quoted uncertainty of $\pm0.2$. The 95\% C.L. is also indicated.
\end{itemize}

From Fig.~\ref{SteelBottleCountsVsHeight-2s-20s-50s} and  Fig.~\ref{SteelBottleStorageOneCommonScalingMCmeas} we conclude that the simulation model reproduces  the distribution of the measured data. The minimal  $\chi^2_\text{red}\approx 4$ in  Fig.~\ref{fig:SteelBottleAlphaSampling} indicates a 2$\sigma$ deviation between simulation and measured data, where $\sigma$ is the square root of the quadratic sum of errors from the measurement and simulation. The relative deviation between measured and simulated data is mostly below 18\%. The deviation at longer  storage times is larger, however, still indicates a close qualitative agreement. We remind here that the simulation model is based on a reduced number of global parameters, see calibrated parameters in Table~\,\ref{table:SimulationParameters}, and thus does not take account of each individual section, the local variation of  surface quality and the exact mechanism of diffuse reflections. Nevertheless, the simulations closely reproduce the measurements.

\begin{table*}
\centering
\begin{tabular}{|c|c c|c c|c c|c c|}
  \hline
Surface  & $V_\text{F}$ & Method &  $\eta$  & Method &  $p_\text{diff}$ & Method &  $\Sigma_\text{atten}$ & Method \\
            & (neV)  &   &  (-)   &   &  (-)  &   &  cm$^{-1}$m/s  &   \\
	\hline
Lid sD$_2$  & 54  & AlMg3 calc.  &  1$\times$10$^{-4}$   & low dep. &  0.10  & low dep.  &  67  & meas.~\cite{Atchison2009} \\
	\hline
Vert. guide  & 220  & NiMo meas. &  5$\times$10$^{-4}$   & gaps calc. &  0.04  & low dep. &  n/a  &  \\
	\hline
Stor. vessel  & 230  & DLC meas. &  11$\times$10$^{-4}$   & meas.+MC &  0.50  & low dep. &  n/a  &  \\
	\hline
Guide South  & 220  & NiMo meas. &  3$\times$10$^{-4}$   & meas.~\cite{Bondar2017} + gaps &  0.05  & meas.+MC &  n/a  &  \\
	\hline
Windows  & 54  & AlMg3 calc.  &  1$\times$10$^{-4}$   & low dep. &  0.10  &  low dep.  &  67  & meas.~\cite{Atchison2009} \\
	\hline
Steel surfaces  & 174  & meas.+MC  &  3.4$\times$10$^{-4}$   & meas.+MC &  0.40  &  meas.+MC  &  n/a  &  \\
	\hline
\end{tabular}
\caption[Simulation parameters]{Parameters of the coatings used in the simulations: optical potential ($V_\text{F}$), loss parameter ($\eta=W/V_\text{F}$), fraction of diffuse (Lambert) reflections ($p_\text{diff}$), and attenuation constant of the material ($\Sigma_\text{atten}$) for 1\,m/s. These numbers represent (i) theoretical values~\cite{Golub1991} (calc.), (ii) measurements discussed in this paper and in~\cite{Bondar2017} (meas.), (iii) simulations benchmarked with measurements in this work (meas. + MC), (iv) geometrical estimations of gaps (gaps.), and (v) rough estimations with a low dependency of the outcome (low dep.).
}
\label{table:SimulationParameters}
\end{table*}

\subsection{Applications of the MCUCN model of the PSI UCN source}

By comparing the MCUCN simulation model to a series of experimental data, we were able to reproduce various measurements within about 2$\sigma$ deviation and to  constrain the loss and diffuse reflection parameters for guides South and West-1 between the source vessel and the beamports, demonstrating that the UCN optics parameters are in a reasonable range. For the energy spectrum of the UCNs generated at the sD$_2$ surface we obtained in this simulation analysis a function $P(E)dE\propto E^{2.7\pm0.2}dE$, which we used instead of the textbook distribution assuming linear dependence. % as  in~\cite{Bison2020}. 
 Further work is planned in order to improve our model of diffuse reflections to better match the experimental data at short flight times.  %These are times during which the UCNs experienced a low number of diffuse reflections, a situation in which the Lambert model of ideally diffuse reflections may be inaccurate.

\begin{figure*}[ht]
\begin{center}
\resizebox{0.50\textwidth}{!}{\includegraphics{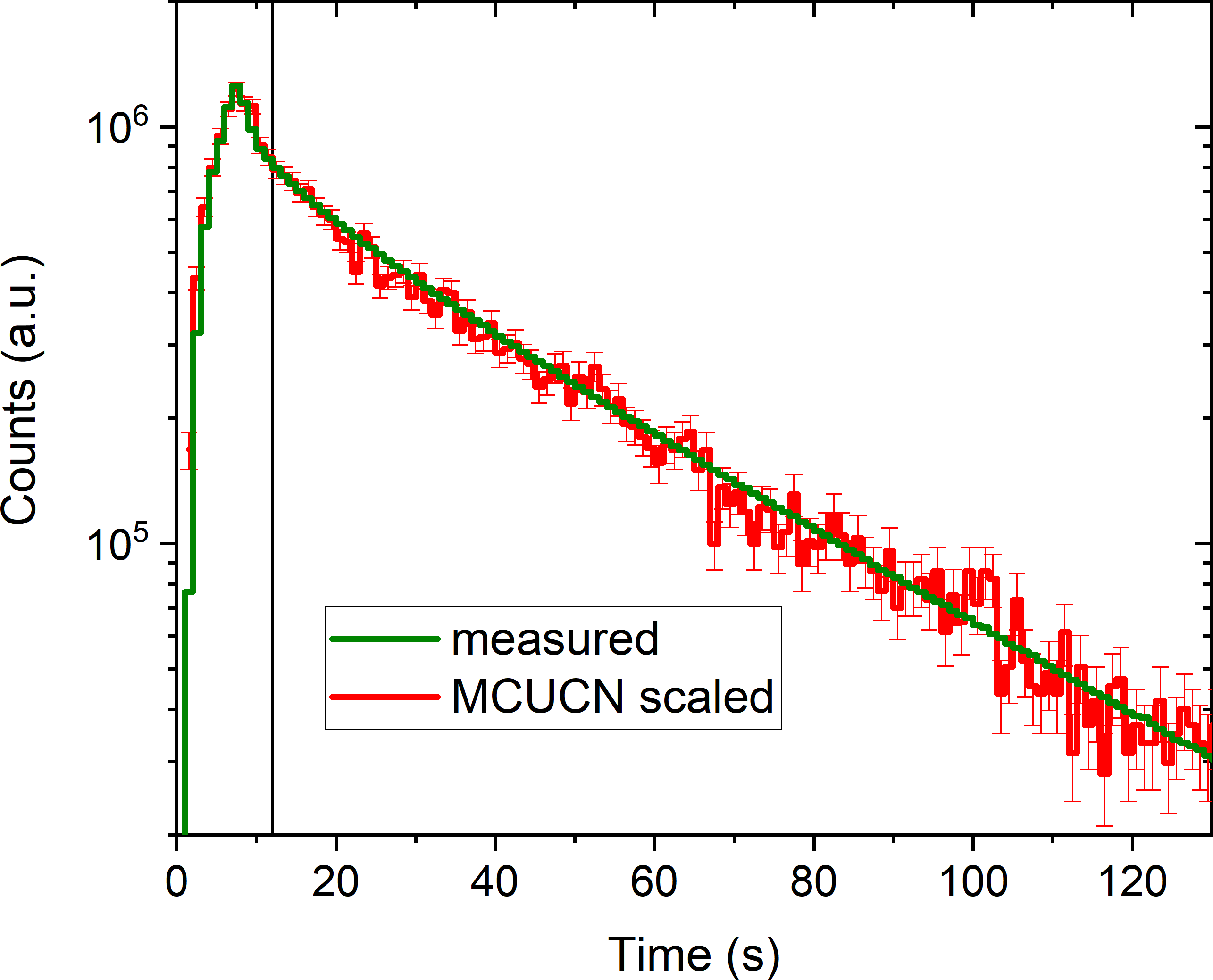}}
\caption{Time spectrum of UCN counts used for the conversion of simulated sampling counts into real UCN counts. The vertical line at 12\,s indicates the lower end of the time interval which was considered to fit the simulation to the data. The proton pulse was on for 8\,s.} 
\label{fig:TOF_MCtoRealConversion}
\end{center}
\end{figure*}

Based on the benchmarks presented in the previous sections, see calibrated parameters in Table~\,\ref{table:SimulationParameters}, the MCUCN model of the PSI UCN source can be used to predict the neutron density in the experiments assuming typical parameters for the experiment volume.  
%as it was done in the case of the n2EDM apparatus ~\cite{Ayres2021n2EDM}.  
To this end, a conversion factor between the number of sampling trajectories in the simulation and the real UCN counts has to be obtained. This can be  done by comparing the simulated and measured arrival-time-spectra at beamport West-1 in direct detection mode, see Fig.\,\ref{fig:TOF_MCtoRealConversion}. The conversion factor is then the  scaling used to match the simulated curve to the measured one. We do this in a range of arrival times in the detector larger than 12\, s, i.e. 4\,s after the end of the proton pulse, once the spectrum is cleaned from neutrons faster than UCNs present mainly during the proton beam pulse. The conversion factor is a function of four input values: (i) the maximal UCN energy generated at the sD$_2$ surface, (ii) the proton pulse length and beam current, (iii) the number of generated trajectories, (iv)  the measured counts at the beamport West-1, which depends on the state of the sD$_2$~\cite{Lauss2021,Anghel2018}. 
The conversion factor obtained at the beamport was found to be compatible with the simulation of the measurements of the UCN density in the standard storage bottle within 15\%. Hence we take this value as the uncertainty on the predictions of the UCN yield at the beamport. 
%This uncertainty is relative to the mean of two values: 1) MC simulation of the vertical extraction (s. section...) which overestimates density by 9\%,  and 2) MC of horizontal extraction which underestimates it by  27\%.
The fit quality in Fig.\,\ref{fig:TOF_MCtoRealConversion} for arrival times in the detector below 12\,s, i.e. for a region with many neutrons above the UCN energy, depends on the parameter of diffuse reflections in the source storage vessel (set here 50\%), which otherwise had no influence on the benchmark procedure when only considering the UCN energy range. This means that faster UCNs that undergo multiple reflections at low angles, have a chance to be directed into the UCN guides, and to propagate further at low reflection angles to the beamport. This arrival-time range below 12 s, however, is not reliably reproduced by the Lambert-model for diffuse reflections.

\begin{figure*}[ht]
\begin{center}
\resizebox{0.50\textwidth}{!}{\includegraphics{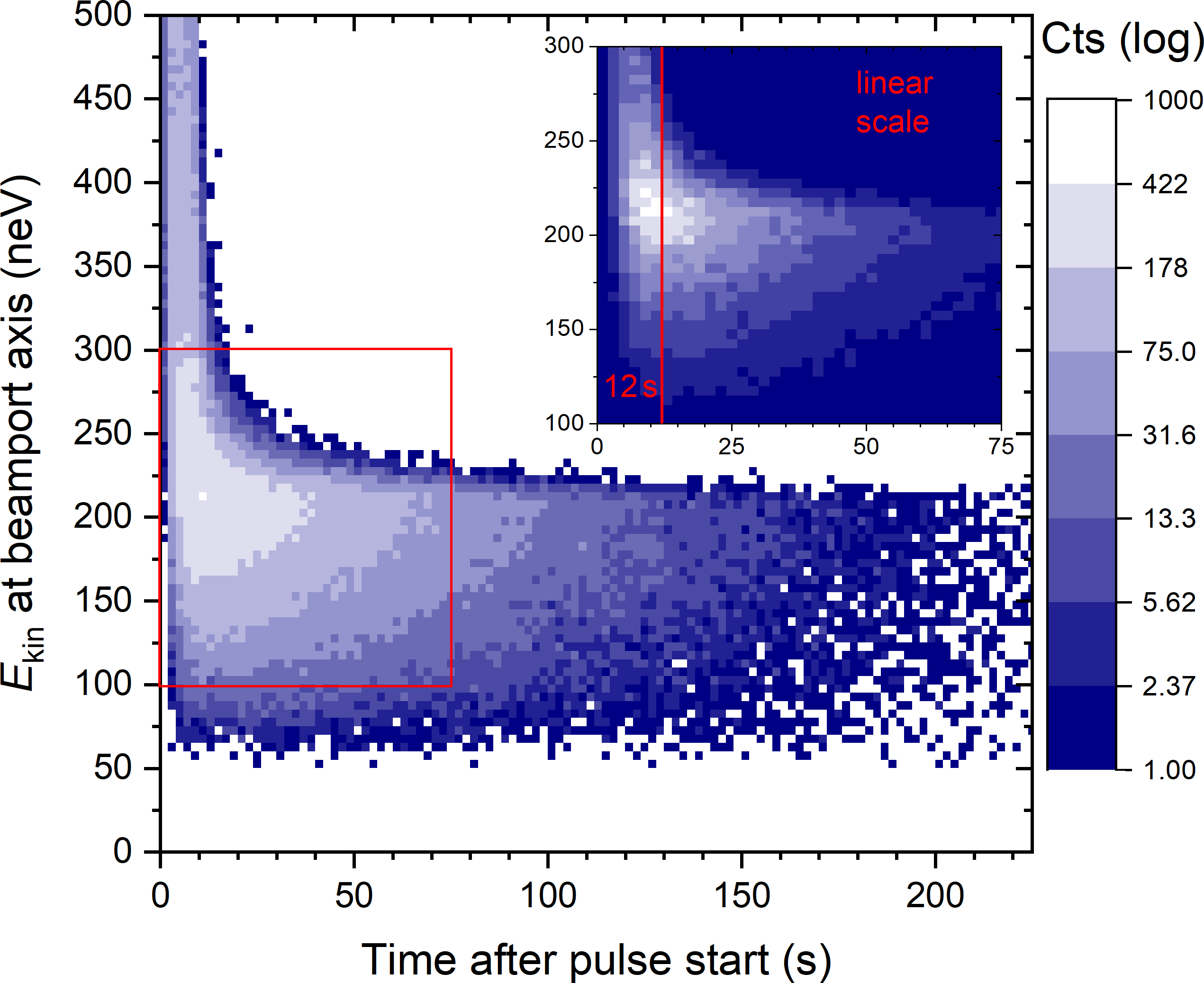}}
\caption{Simulated kinetic energy of UCNs at the beamport versus arrival time in the detector.
A cleaning effect removing higher energy neutrons is visible. The insert is a zoom-in to the region with the highest counts displayed in a linear scale. Two regions can be discerned: times below 12\,s when the fastest neutrons can arrive, and times after 12\,s (4\,s after the 8\,s long proton pulse) when the spectrum almost only consists of UCNs.} 
\label{fig:spectrum_cleaning_after_12s}
\end{center}
\end{figure*}

\begin{figure*}[ht]
\begin{center}
\resizebox{0.50\textwidth}{!}{\includegraphics{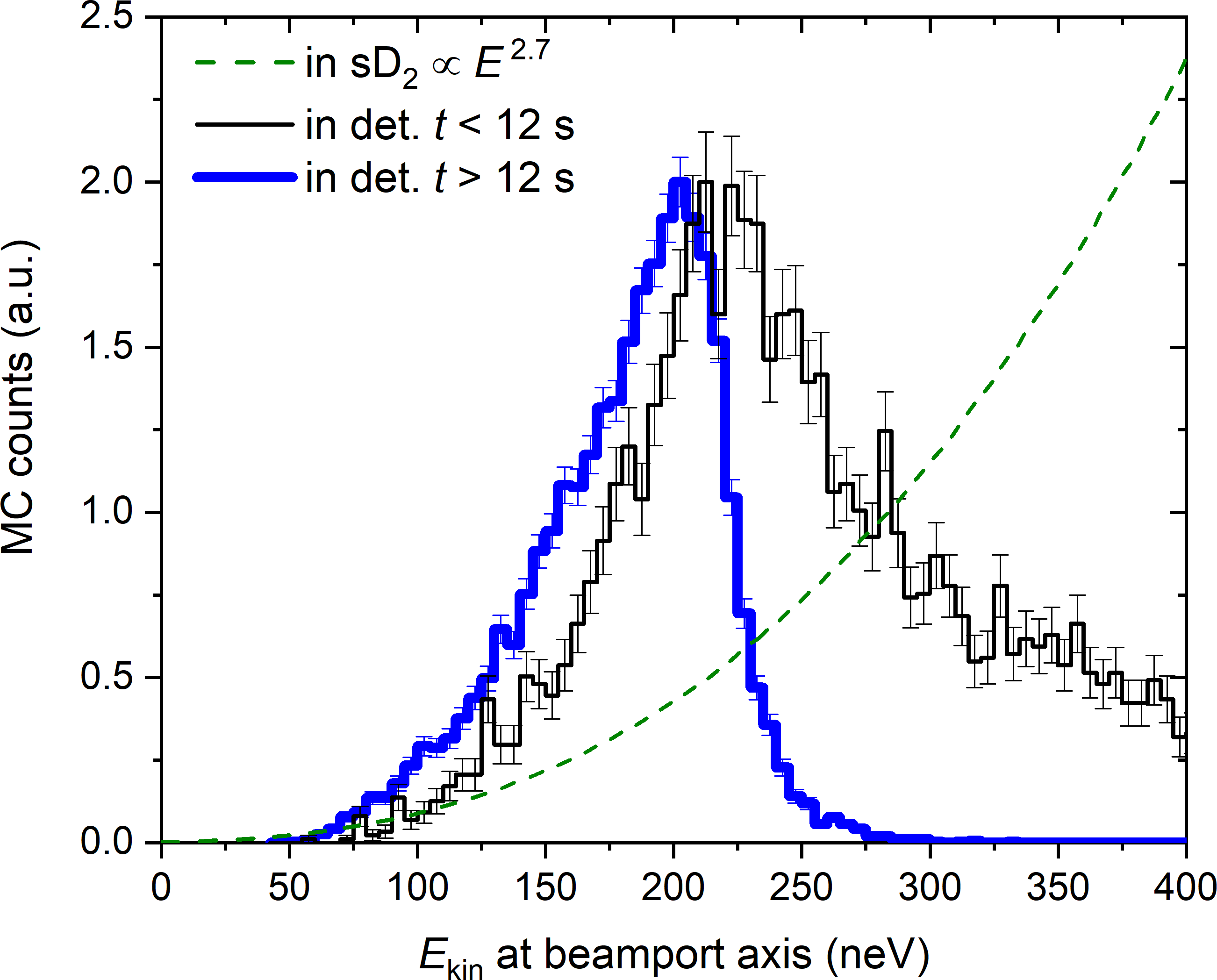}}
\caption{The kinetic energy spectra of UCNs calculated after the detector window. Black line: maximal arrival times of 12\,s. Blue line: minimal arrival times of 12\,s. Green dashed line: case of perfect transmission i.e. the spectral profile of neutrons exiting the sD$_2$ moderator.} 
\label{fig:SpectrumProfilesInSD2_inDetector}
\end{center}
\end{figure*}

The spectrum is cleaned from neutrons above the UCN energy range about 12\,s after the start of the proton pulse, as shown in  Fig.\,\ref{fig:spectrum_cleaning_after_12s}. 
The spectrum profiles at the beamport for UCNs arriving before 12\,s and after 12\,s in the detector are displayed in Fig.\,\ref{fig:SpectrumProfilesInSD2_inDetector}. The kinetic energy was calculated at the level of the beam axis. The green dashed line represents the energy spectrum generated at the sD$_2$ surface, and was shifted according to the height of the beamport.

\begin{figure*}[ht]
\begin{center}
\resizebox{0.50\textwidth}{!}{\includegraphics{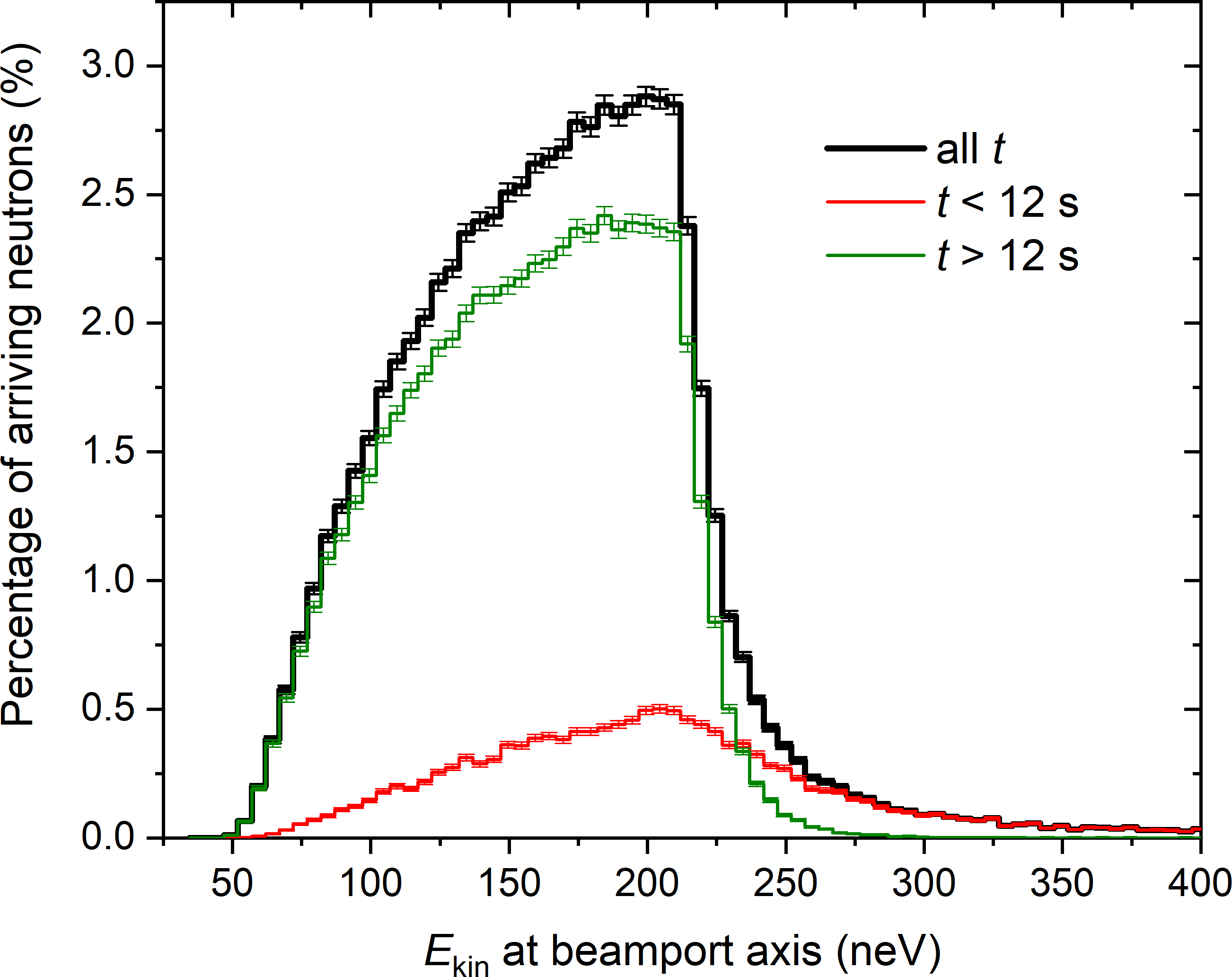}}
\caption{Simulated percentage of neutrons starting at the sD$_2$ arriving in the detector (including the detector window) as a function of kinetic energy at the height of the beam axis. The black line represents all the transmitted neutrons independent of the arrival time. The red and green lines represent the fractions arriving before and after 12\,s in the detector.} 
\label{fig:TransmissionFromSD2toDetectionTiming}
\end{center}
\end{figure*}

It can also be useful for the analysis of the UCN source output to calculate the transmission of UCNs between the sD$_2$ moderator and the detector as a function of the kinetic energy. Figure\,\ref{fig:TransmissionFromSD2toDetectionTiming} shows the simulated energy dependence of the UCN transmission between the sD$_2$ surface and detection position just after the detector window. The black solid line is the transmission of the UCNs in the entire 300\,s period between the pulses, $N/N_0$, where $N$ is the ratio of the counts integrated over all arrival times and, $N_0$ is the number of UCNs escaping through the surface of the sD$_2$.
In order to only measure neutrons in the UCN energy range, it can be useful in the data-analysis to cut the time spectrum in the detector at 12\,s. Therefore, in the simulation it is reasonable to calculate the part of the transmission that is only related to the UCNs detected 12\,s after the start of the pulse, $N_{t>12\,s}/N_0$. This is represented by the green line in Fig.\,\ref{fig:TransmissionFromSD2toDetectionTiming}.
The black solid line indicates that the maximal transmission, 3\,\% is reached around 200 neV, a 25\% lower value  than previously obtained~\cite{Bison2020}, when an effective fraction of diffuse reflections of 0.02  was assumed in the source guides. 
Table~\ref{table:LossesInComponents} summarizes the location of UCN losses during the 300\,s period between the pulses, including  the fraction of decayed and detected neutrons. For various components of the UCN optics, we counted the events in the competing loss channels:  absorption or up-scattering at reflection or transmission, and beta decay. This number is divided by $N_0$. The largest part of the UCNs lost is identified during transmission through the lid of the moderator vessel (counted in both flight directions), which is indispensable for safety reasons. The second largest loss is due to UCNs returning to the sD$_2$ during the  8\,s pulse, when the storage vessel flaps must be open. The third largest loss fraction is in the vertical guide between the lid and the central storage vessel flaps (including the support structure), see Fig.~\ref{fig:ping-pong-setup}, caused by exceeding the optical potential barrier at large emission angles.
%Increasing the optical potential in the vertical guide from 220 neV (NiMo) to  340 neV ($^{58}$Ni) would reduce this loss component by a factor of 28, however it would increase the counts in the detector only by about 8\%, since the largest fraction of the ``recovered'' UCNs is lost in the other components. 

%The comparison of the simulated spectrum to direct measurements is the subject of a next study. These direct spectrum measurements are performed using a chopper and an oscillating detector~\cite{Rozpedzik2019}. Preliminary results suggest that in the first type of measurements the velocity peak position is higher by~$\sim0.5$m/s, and in the second the velocity peak position matches well the full source simulation. The analysis of these measurements is still ongoing and will be published later.

\begin{table*}
\centering
\begin{tabular}{|c|c|c|}
  \hline
Component & $N_\text{loss}/N_0$  (\%) & $N_\text{loss}/N_0$   (\%) \\
                 & 200 neV  &   50 $-$ 300 neV   \\
	\hline
Moderator (returned neutrons)  &  19.97 $\pm$ 0.14  & 16.63 $\pm$ 0.13 \\
	\hline
Moderator vessel  lid &  57.38 $\pm$ 0.24   & 52.46 $\pm$ 0.23  \\
	\hline
Vertical guide  &  12.35 $\pm$ 0.11  &  17.24 $\pm$ 0.13  \\
	\hline
Source storage vessel &  4.44 $\pm$ 0.07   &  10.71 $\pm$ 0.10  \\
	\hline
West-1  guides &   0.82 $\pm$ 0.03  &  0.93 $\pm$ 0.03  \\
	\hline
Vacuum safety window West-1 &  1.14 $\pm$ 0.03   & 0.48 $\pm$ 0.02  \\
	\hline
Detector foil West-1 &  0.62 $\pm$ 0.02   &  0.25 $\pm$ 0.02 \\
	\hline
All losses via West-2  &  0.00 $\pm$ 0.00  &  0.05 $\pm$ 0.01 \\
	\hline
Decayed  &  0.44 $\pm$ 0.02   & 0.20 $\pm$ 0.01 \\
	\hline
Detected in West-1  &  2.84 $\pm$ 0.05   &  1.04 $\pm$ 0.03 \\
	\hline
\end{tabular}
\caption[Losses in components]{Localization of UCN losses in the various components of the MCUCN model.  The fraction of the UCNs lost over a cycle was calculated for 200 neV at the beam axis and for the interval 50 $-$ 300 neV (compare to spectra in 
Fig.~\ref{fig:SpectrumProfilesInSD2_inDetector}) relative to the initial number of UCNs, $N_0$, started at the sD$_2$ surface during the pulse. The guide shutter South was closed, and the  guide shutters West-1 and West-2 were open.}
\label{table:LossesInComponents}
\end{table*}

%%%%%%%%%%%%%%%%%%%%%%%%%%%%%%%%%%%
\section{Conclusions}
\label{conclusions}

Our UCN transport and storage measurements, and their reproductions in simulations presented here constitute an important part of a series of studies aiming at the characterization of the UCN source components at PSI in view of further improvement of their performance.

Measurements of the UCN transmission from beamport-to-beamport through the guide system and central storage vessel of the PSI UCN source were performed with high statistics.  The timing was chosen such that signal and background regions were well separated. 
The transmitted fraction of UCNs agrees well with detailed Monte Carlo simulations, which were benchmarked along with additional measurements.

A second type of experiment, UCN storage at different heights above the beamport in a standard storage  bottle, allowed maximizing the UCN density in this storage bottle by optimizing its vertical position. The UCN density shows a flat maximum at around 800\,mm of height.
This optimization procedure enabled estimating an approximate function for the energy spectrum of the UCNs exiting the moderator. This obtained function indicates a loss of neutrons in the lowest energy range, as expected, considering the poly-crystalline structure of the sD$_2$ moderator~\cite{Atchison2005b,Brys2007}.

By comparing the MCUCN simulation model to experimental data, we were able to reproduce various measurements and extract UCN loss and diffuse reflection 
parameters for the South and West-1 source guides.  This demonstrates that the UCN optics of the PSI UCN source is well understood.  The present study provides a guideline for possible  improvements of source components.

We could also derive the energy spectrum of UCNs arriving at the detector at the beamport. It is clearly shown that UCNs with energies well above the neutron optical potential of the UCN guides arrive at the beamport despite of the UCN guide length of about 8\,m. We have shown that these neutrons disappear within about 12\,s and a storable UCN energy spectrum remains.

As main application, the benchmarked MCUCN model presented here was employed to predict the statistical sensitivity of  the n2EDM apparatus connected to the UCN source model and to find  the optimal height of the double-chamber, assuming typical parameters for the UCN optics of the experiment~\cite{Ayres2021n2EDM}. Our simulation model will be used to support the planning and analysis of future experiments at the UCN source.

%%%%%%%%%%%%%%%%%%%%%%%%%%%%%%%%%%%
\section*{Acknowledgements}

Most of the measurements were performed in the context of the PhD thesis work of one of the co-authors, D.~Ries~\cite{Ries2016}. 
We would like to thank the many people who supported in various ways the presented work: 
L.~Goeltl and S.~Komposch contributed in the early phase of the measurement. 
We acknowledge 
the PSI proton accelerator operations section, 
all colleagues who have been contributing to the UCN source construction 
and operation at PSI, 
and especially the BSQ group which has been operating the PSI UCN source during the measurements, 
namely B.~Blau, P.~Erisman and also A.~Anghel. 
Excellent technical support by F.~Burri and M.~Meier is acknowledged.
This work was supported by the Swiss National Science Foundation
Projects 200020\_137664, 200020\_149813, 200020\_163413, 200021\_117696 and 200021\_178951. 
Special thanks for granting access to the computing facility PL-Grid~\cite{PLGrid}.

\bibliographystyle{unsrt}
\bibliography{PPandSB}

\begin{thebibliography}{10}

\bibitem{Golub1991}
R.~Golub, D.J. Richardson, and S.K. Lamoreaux.
\newblock {\em Ultra-Cold Neutrons}.
\newblock Adam Hilger, Bristol, Philadelphia, and New York, 1991.

\bibitem{Baker2006}
C.~A. Baker, D.~D. Doyle, P.~Geltenbort, K.~Green, M.~G.~D. van~der Grinten,
  P.~G. Harris, P.~Iaydjiev, S.~N. Ivanov, D.~J.~R. May, J.~M. Pendlebury,
  J.~D. Richardson, D.~Shiers, and K.~F. Smith.
\newblock Improved experimental limit on the electric dipole moment of the
  neutron.
\newblock {\em Phys. Rev. Lett.}, 97(13):131801, September 2006.

\bibitem{Baker2011}
C.A. Baker, G.~Ban, K.~Bodek, M.~Burghoff, Z.~Chowdhuri, M.~Daum, M.~Fertl,
  B.~Franke, P.~Geltenbort, K.~Green, M.G.D. van~der Grinten, E.~Gutsmiedl,
  P.G. Harris, R.~Henneck, P.~Iaydjiev, S.N. Ivanov, N.~Khomutov, M.~Kasprzak,
  K.~Kirch, S.~Kistryn, S.~Knappe-Gr{\"u}neberg, A.~Knecht, P.~Knowles,
  A.~Kozela, B.~Lauss, T.~Lefort, Y.~Lemiere, O.~Naviliat-Cuncic, J.M.
  Pendlebury, E.~Pierre, F.M. Piegsa, G.~Pignol, G.~Quemener, S.~Roccia,
  P.~Schmidt-Wellenburg, D.~Shiers, K.F. Smith, A.~Schnabel, L.~Trahms,
  A.~Weis, J.~Zejma, J.~Zenner, and G.~Zsigmond.
\newblock The search for the neutron electric dipole moment at the {Paul
  Scherrer Institute}.
\newblock {\em Physics Proc.}, 17(0):159 -- 167, 2011.
\newblock 2nd International Workshop on the Physics of fundamental Symmetries
  and Interactions - PSI2010.

\bibitem{SerebrovEDM2015}
A.~P. Serebrov, E.~A. Kolomenskiy, A.~N. Pirozhkov, I.~A. Krasnoschekova, A.~V.
  Vassiljev, A.~O. Polyushkin, M.~S. Lasakov, A.~N. Murashkin, V.~A. Solovey,
  A.~K. Fomin, I.~V. Shoka, O.~M. Zherebtsov, P.~Geltenbort, S.~N. Ivanov,
  O.~Zimmer, E.~B. Alexandrov, S.~P. Dmitriev, and N.~A. Dovator.
\newblock New search for the neutron electric dipole moment with ultracold
  neutrons at {ILL}.
\newblock {\em Phys. Rev. C}, 92:055501, Nov 2015.

\bibitem{Pendlebury2015}
J.~M. Pendlebury, S.~Afach, N.~J. Ayres, C.~A. Baker, G.~Ban, G.~Bison,
  K.~Bodek, M.~Burghoff, P.~Geltenbort, K.~Green, W.~C. Griffith, M.~van~der
  Grinten, Z.~D. Gruji\ifmmode~\acute{c}\else \'{c}\fi{}, P.~G. Harris,
  V.~H\'elaine, P.~Iaydjiev, S.~N. Ivanov, M.~Kasprzak, Y.~Kermaidic, K.~Kirch,
  H.-C. Koch, S.~Komposch, A.~Kozela, J.~Krempel, B.~Lauss, T.~Lefort,
  Y.~Lemi\`ere, D.~J.~R. May, M.~Musgrave, O.~Naviliat-Cuncic, F.~M. Piegsa,
  G.~Pignol, P.~N. Prashanth, G.~Qu\'em\'ener, M.~Rawlik, D.~Rebreyend, J.~D.
  Richardson, D.~Ries, S.~Roccia, D.~Rozpedzik, A.~Schnabel,
  P.~Schmidt-Wellenburg, N.~Severijns, D.~Shiers, J.~A. Thorne, A.~Weis, O.~J.
  Winston, E.~Wursten, J.~Zejma, and G.~Zsigmond.
\newblock Revised experimental upper limit on the electric dipole moment of the
  neutron.
\newblock {\em Phys. Rev. D}, 92:092003, Nov 2015.

\bibitem{Ito2018}
T.~M. Ito, E.~R. Adamek, N.~B. Callahan, J.~H. Choi, S.~M. Clayton,
  C.~Cude-Woods, S.~Currie, X.~Ding, D.~E. Fellers, P.~Geltenbort, S.~K.
  Lamoreaux, C.-Y. Liu, S.~MacDonald, M.~Makela, C.~L. Morris, R.~W. Pattie,
  J.~C. Ramsey, D.~J. Salvat, A.~Saunders, E.~I. Sharapov, S.~Sjue, A.~P.
  Sprow, Z.~Tang, H.~L. Weaver, W.~Wei, and A.~R. Young.
\newblock Performance of the upgraded ultracold neutron source at {Los Alamos
  National Laboratory} and its implication for a possible neutron electric
  dipole moment experiment.
\newblock {\em Phys. Rev. C}, 97:012501, Jan 2018.

\bibitem{Ahmed2019}
M.W. Ahmed, R.~Alarcon, A.~Aleksandrova, S.~Bae{\ss}ler, L.~Barron-Palos, L.M.
  Bartoszek, D.H. Beck, M.~Behzadipour, I.~Berkutov, J.~Bessuille, M.~Blatnik,
  M.~Broering, L.J. Broussard, M.~Busch, R.~Carr, V.~Cianciolo, S.M. Clayton,
  M.D. Cooper, C.~Crawford, S.A. Currie, C.~Daurer, R.~Dipert, K.~Dow,
  D.~Dutta, Y.~Efremenko, C.B. Erickson, B.W. Filippone, N.~Fomin, H.~Gao,
  R.~Golub, C.R. Gould, G.~Greene, D.G. Haase, D.~Hasell, A.I. Hawari, M.E.
  Hayden, A.~Holley, R.J. Holt, P.R. Huffman, E.~Ihloff, S.K. Imam, T.M. Ito,
  M.~Karcz, J.~Kelsey, D.P. Kendellen, Y.J. Kim, E.~Korobkina, W.~Korsch, S.K.
  Lamoreaux, E.~Leggett, K.K.H. Leung, A.~Lipman, C.Y. Liu, J.~Long, S.W.T.
  MacDonald, M.~Makela, A.~Matlashov, J.D. Maxwell, M.~Mendenhall, H.O. Meyer,
  R.G. Milner, P.E. Mueller, N.~Nouri, C.M. O{\textquotesingle}Shaughnessy,
  C.~Osthelder, J.C. Peng, S.I. Penttila, N.S. Phan, B.~Plaster, J.C. Ramsey,
  T.M. Rao, R.P. Redwine, A.~Reid, A.~Saftah, G.M. Seidel, I.~Silvera,
  S.~Slutsky, E.~Smith, W.M. Snow, W.~Sondheim, S.~Sosothikul, T.D.S.
  Stanislaus, X.~Sun, C.M. Swank, Z.~Tang, R.~Tavakoli Dinani, E.~Tsentalovich,
  C.~Vidal, W.~Wei, C.R. White, S.E. Williamson, L.~Yang, W.~Yao, and A.R.
  Young.
\newblock A new cryogenic apparatus to search for the neutron electric dipole
  moment.
\newblock {\em Journal of Instrumentation}, 14(11):P11017--P11017, {N}ov 2019.

\bibitem{Wurm2019}
D.~Wurm, Douglas~H. Beck, T.~Chupp, S.~Degenkolb, K.~Fierlinger, P.~Fierlinger,
  H.~Filter, S.~Ivanov, Ch. Klau, M.~Kreuz, E.~{Leli\`evre-Berna}, T.~Lins,
  J.~{Meichelb\"ock}, Th. Neulinger, R.~Paddock, F.~{R\"ohrer}, M.~Rosner,
  A.~P. Serebrov, J.~Taggart Singh, R.~Stoepler, S.~Stuiber, M.~Sturm,
  B.~Taubenheim, X.~Tonon, M.~Tucker, {M. van der} Grinten, and O.~Zimmer.
\newblock The {PanEDM} neutron electric dipole moment experiment at the {ILL}.
\newblock {\em EPJ Web Conf.}, 219:02006, 2019.

\bibitem{Martin2020}
J.W. Martin.
\newblock Current status of neutron electric dipole moment experiments.
\newblock {\em Journal of Physics: Conference Series}, 1643:012002, {D}ec 2020.

\bibitem{Abel2020}
C.~Abel, S.~Afach, N.~J. Ayres, C.~A. Baker, G.~Ban, G.~Bison, K.~Bodek,
  V.~Bondar, M.~Burghoff, E.~Chanel, Z.~Chowdhuri, P.-J. Chiu, B.~Clement,
  C.~B. Crawford, M.~Daum, S.~Emmenegger, L.~Ferraris-Bouchez, M.~Fertl,
  P.~Flaux, B.~Franke, A.~Fratangelo, P.~Geltenbort, K.~Green, W.~C. Griffith,
  M.~van~der Grinten, Z.~D. Gruji\ifmmode~\acute{c}\else \'{c}\fi{}, P.~G.
  Harris, L.~Hayen, W.~Heil, R.~Henneck, V.~H\'elaine, N.~Hild, Z.~Hodge,
  M.~Horras, P.~Iaydjiev, S.~N. Ivanov, M.~Kasprzak, Y.~Kermaidic, K.~Kirch,
  A.~Knecht, P.~Knowles, H.-C. Koch, P.~A. Koss, S.~Komposch, A.~Kozela,
  A.~Kraft, J.~Krempel, M.~Ku\ifmmode~\acute{z}\else \'{z}\fi{}niak, B.~Lauss,
  T.~Lefort, Y.~Lemi\`ere, A.~Leredde, P.~Mohanmurthy, A.~Mtchedlishvili,
  M.~Musgrave, O.~Naviliat-Cuncic, D.~Pais, F.~M. Piegsa, E.~Pierre, G.~Pignol,
  C.~Plonka-Spehr, P.~N. Prashanth, G.~Qu\'em\'ener, M.~Rawlik, D.~Rebreyend,
  I.~Rien\"acker, D.~Ries, S.~Roccia, G.~Rogel, D.~Rozpedzik, A.~Schnabel,
  P.~Schmidt-Wellenburg, N.~Severijns, D.~Shiers, R.~Tavakoli~Dinani, J.~A.
  Thorne, R.~Virot, J.~Voigt, A.~Weis, E.~Wursten, G.~Wyszynski, J.~Zejma,
  J.~Zenner, and G.~Zsigmond.
\newblock Measurement of the permanent electric dipole moment of the neutron.
\newblock {\em Phys. Rev. Lett.}, 124:081803, Feb 2020.

\bibitem{Bison2017}
G.~Bison, M.~Daum, K.~Kirch, B.~Lauss, D.~Ries, P.~Schmidt-Wellenburg,
  G.~Zsigmond, T.~Brenner, P.~Geltenbort, T.~Jenke, O.~Zimmer, M.~Beck,
  W.~Heil, J.~Kahlenberg, J.~Karch, K.~Ross, K.~Eberhardt, C.~Geppert,
  S.~Karpuk, T.~Reich, C.~Siemensen, Y.~Sobolev, and N.~Trautmann.
\newblock Comparison of ultracold neutron sources for fundamental physics
  measurements.
\newblock {\em Phys. Rev. C}, 95:045503, Apr 2017.

\bibitem{Leung2019source}
K.~K.~H. Leung, G.~Muhrer, T.~Hügle, T.~M. Ito, E.~M. Lutz, M.~Makela, C.~L.
  Morris, R.~W. Pattie, A.~Saunders, and A.~R. Young.
\newblock A next-generation inverse-geometry spallation-driven ultracold
  neutron source.
\newblock {\em Journal of Applied Physics}, 126(22):224901, 2019.

\bibitem{Schreyer2020}
W.~Schreyer, C.A. Davis, S.~Kawasaki, T.~Kikawa, C.~Marshall, K.~Mishima,
  T.~Okamura, and R.~Picker.
\newblock Optimizing neutron moderators for a spallation-driven
  ultracold-neutron source at {TRIUMF}.
\newblock {\em Nuclear Instruments and Methods in Physics Research Section A:
  Accelerators, Spectrometers, Detectors and Associated Equipment}, 959:163525,
  2020.

\bibitem{Lauss2014}
B.~Lauss.
\newblock {Ultracold Neutron Production at the Second Spallation Target of the
  Paul Scherrer Institute}.
\newblock {\em Phys. Proc.}, 51:98, 2014.

\bibitem{Becker2015}
H.~Becker, G.~Bison, B.~Blau, Z.~Chowdhuri, J.~Eikenberg, M.~Fertl, K.~Kirch,
  B.~Lauss, G.~Perret, D.~Reggiani, D.~Ries, P.~Schmidt-Wellenburg, V.~Talanov,
  M.~Wohlmuther, and G.~Zsigmond.
\newblock Neutron production and thermal moderation at the {PSI UCN} source.
\newblock {\em Nucl. Instrum. Methods A}, 777(0):20--27, 2015.

\bibitem{Bison2020}
G.~Bison, B.~Blau, M.~Daum, L.~G{\"o}ltl, R.~Henneck, K.~Kirch, B.~Lauss,
  D.~Ries, P.~Schmidt-Wellenburg, and G.~Zsigmond.
\newblock Neutron optics of the {PSI} ultracold-neutron source:
  characterization and simulation.
\newblock {\em The European Physical Journal A}, 56(2):33, Feb 2020.

\bibitem{Lauss2021}
B.~Lauss and B.~Blau.
\newblock {UCN, the ultracold neutron source -- neutrons for particle physics}.
\newblock {\em SciPost Phys. Proc.}, 5:004, 2021.

\bibitem{Wohlmuther2006}
M.~Wohlmuther and G.~Heidenreich.
\newblock {The spallation target of the ultra-cold neutron source UCN at PSI}.
\newblock {\em Nucl. Instrum. Methods A}, 564:51, 2006.

\bibitem{Atchison2009a}
F.~Atchison, B.~Blau, K.~Bodek, B.~{van den Brandt}, T.~Bryś, M.~Daum,
  P.~Fierlinger, A.~Frei, P.~Geltenbort, P.~Hautle, R.~Henneck, S.~Heule,
  A.~Holley, M.~Kasprzak, K.~Kirch, A.~Knecht, J.A. Konter, M.~Kuźniak, C.-Y.
  Liu, C.L. Morris, A.~Pichlmaier, C.~Plonka, Y.~Pokotilovski, A.~Saunders,
  Y.~Shin, D.~Tortorella, M.~Wohlmuther, A.R. Young, J.~Zejma, and G.~Zsigmond.
\newblock Investigation of solid {D2}, {O2} and {CD4} for ultracold neutron
  production.
\newblock {\em Nuclear Instruments and Methods in Physics Research Section A:
  Accelerators, Spectrometers, Detectors and Associated Equipment},
  611(2):252--255, 2009.
\newblock Particle Physics with Slow Neutrons.

\bibitem{Atchison2011}
F.~Atchison, B.~Blau, K.~Bodek, B.~van~den Brandt, T.~Bry{\'{s}}, M.~Daum,
  P.~Fierlinger, P.~Geltenbort, P.~Hautle, R.~Henneck, S.~Heule, A.~Holley,
  M.~Kasprzak, K.~Kirch, A.~Knecht, J.~A. Konter, M.~Ku{\'{z}}niak, C.-Y. Liu,
  A.~Pichlmaier, C.~Plonka, Y.~Pokotilovski, A.~Saunders, D.~Tortorella,
  M.~Wohlmuther, A.~R. Young, J.~Zejma, and G.~Zsigmond.
\newblock Production of ultracold neutrons from cryogenic {D2}, {O2}, and
  {C2H4} converters.
\newblock {\em {EPL} (Europhysics Letters)}, 95(1):12001, June 2011.

\bibitem{Anghel2018}
A.~Anghel, T.~L. Bailey, G.~Bison, B.~Blau, L.~J. Broussard, S.~M. Clayton,
  C.~Cude-Woods, M.~Daum, A.~Hawari, N.~Hild, P.~Huffman, T.~M. Ito, K.~Kirch,
  E.~Korobkina, B.~Lauss, K.~Leung, E.~M. Lutz, M.~Makela, G.~Medlin, C.~L.
  Morris, R.~W. Pattie, D.~Ries, A.~Saunders, P.~Schmidt-Wellenburg,
  V.~Talanov, A.~R. Young, B.~Wehring, C.~White, M.~Wohlmuther, and
  G.~Zsigmond.
\newblock Solid deuterium surface degradation at ultracold neutron sources.
\newblock {\em The European Physical Journal A}, 54(9):148, Sep 2018.

\bibitem{Blau2016}
B.~Blau, M.~Daum, M.~Fertl, P.~Geltenbort, L.~Goeltl, R.~Henneck, K.~Kirch,
  A.~Knecht, B.~Lauss, P.~Schmidt-Wellenburg, and G.~Zsigmond.
\newblock A prestorage method to measure neutron transmission of ultracold
  neutron guides.
\newblock {\em Nucl. Instrum. Methods A}, 807:30 -- 40, 2016.

\bibitem{Goeltl2012}
L.~G{\"o}ltl.
\newblock {\em {Characterization of the {PSI} ultra-cold neutron source}}.
\newblock PhD thesis, ETH Z{\"u}rich, No.20350, 2012.

\bibitem{Cascade2013}
Martin Klein.
\newblock {\em CDT Detector Control - Reference Manual}.
\newblock CD-T Detector Technologies GmbH, Hans Bunte Strasse 8-10, 69123
  Heidelberg, Germany, {http://n-cdt.com}, April 2013.

\bibitem{Bison2016}
G.~Bison, F.~Burri, M.~Daum, K.~Kirch, J.~Krempel, B.~Lauss, M.~Meiter,
  D.~Ries, P.~Schmidt-Wellenburg, and G.~Zsigmond.
\newblock An ultracold neutron storage bottle for {UCN} density measurements.
\newblock {\em Nucl. Instrum. Methods A}, 830:449, 2016.

\bibitem{Zsigmond2018}
G.~Zsigmond.
\newblock The {MCUCN} simulation code for ultracold neutron physics.
\newblock {\em Nucl. Instrum. Methods A}, 881:16--26, 2018.

\bibitem{Bondar2017}
V.~Bondar, S.~Chesnevskaya, M.~Daum, B.~Franke, P.~Geltenbort, L.~Goeltl,
  E.~Gutsmiedl, J.~Karch, M.~Kasprzak, G.~Kessler, K.~Kirch, H.-C. Koch,
  A.~Kraft, T.~Lauer, B.~Lauss, E.~Pierre, G.~Pignol, D.~Reggiani,
  P.~Schmidt-Wellenburg, Yu. Sobolev, T.~Zechlau, and G.~Zsigmond.
\newblock Losses and depolarisation of ultracold neutrons on neutron guide and
  storage materials.
\newblock {\em Phys. Rev. C}, 96:035205, 2017.

\bibitem{Atchison2009}
F.~Atchison, Blau B., A.~Bollhalder, M.~Daum, P.~Fierlinger, P.~Geltenbort, and
  G.~Hampel.
\newblock Transmission of very slow neutrons through material foils and its
  influence on the design of ultracold neutron sources.
\newblock {\em Nucl. Instrum. Methods A}, A 608:144--151, 2009.

\bibitem{Ries2016}
D.~Ries.
\newblock {\em {Characterisation and Optimisation of the Source for Ultracold
  Neutrons at the Paul Scherrer Institute}}.
\newblock PhD thesis, ETH Z{\"u}rich, No.23671, 2016.

\bibitem{Ignatovich1990}
V.K. Ignatovich.
\newblock {\em The Physics of Ultracold Neutrons}.
\newblock Clarendon, Oxford, 1990.

\bibitem{ROGERS1975}
D.W.O. Rogers.
\newblock Analytic and graphical methods for assigning errors to parameters in
  non-linear least squares fitting.
\newblock {\em Nuclear Instruments and Methods}, 127(2):253--260, 1975.

\bibitem{Atchison2005b}
F.~Atchison, B.~Blau, B.~van~den Brandt, Brys T., M.~Daum, P.~Fierlinger,
  P.~Hautle, R.~Henneck, S.~Heule, M.~Kasprzak, K.~Kirch, J.~Kohlbrecher,
  G.~Kuehen, J.~A. Konter, A.~Pichlmaier, A.~Wokaun, K.~Bodek, P.~Geltenbort,
  and J.~Zmeskal.
\newblock Measured total cross sections of slow neutrons scattered by solid
  deuterium and implications for ultracold neutron sources.
\newblock {\em Phys. Rev. Lett.}, 95:182502, 2005.

\bibitem{Brys2007}
T.~Brys.
\newblock {\em Extraction of ultracold neutrons from a solid deuterium source}.
\newblock PhD thesis, ETH Z{\"u}rich, No. 17350, 2007.

\bibitem{Ayres2021n2EDM}
N.~J. Ayres, G.~Ban, L.~Bienstman, G.~Bison, K.~Bodek, V.~Bondar, T.~Bouillaud,
  E.~Chanel, J.~Chen, P.~J. Chiu, B.~Clément, C.~Crawford, M.~Daum,
  B.~Dechenaux, C.~B. Doorenbos, S.~Emmenegger, L.~Ferraris-Bouchez, M.~Fertl,
  A.~Fratangelo, P.~Flaux, D.~Goupillière, W.~C. Griffith, Z.~D. Grujic, P.~G.
  Harris, K.~Kirch, P.~A. Koss, J.~Krempel, B.~Lauss, T.~Lefort, Y.~Lemière,
  A.~Leredde, M.~Meier, J.~Menu, D.~A. Mullins, O.~Naviliat-Cuncic, D.~Pais,
  F.~M. Piegsa, G.~Pignol, G.~Quéméner, M.~Rawlik, D.~Rebreyend,
  I.~Rienäcker, D.~Ries, S.~Roccia, K.~U. Ross, D.~Rozpedzik, W.~Saenz,
  P.~Schmidt-Wellenburg, A.~Schnabel, N.~Severijns, B.~Shen, T.~Stapf,
  K.~Svirina, R.~Tavakoli Dinani, S.~Touati, J.~Thorne, R.~Virot, J.~Voigt,
  N.~Yazdandoost, J.~Zejma, and G.~Zsigmond.
\newblock The design of the {n2EDM} experiment.
\newblock {\em Eur. Phys. J. C}, 81:512, 2021.

\bibitem{PLGrid}
Polish {Grid} {Infrastructure} {PL-Grid}, http://www.plgrid.pl/en.

\end{thebibliography}

\clearpage

\end{document}